\newcommand\apjcls{1}
\newcommand\aastexcls{2}
\newcommand\othercls{3}

\newcommand\papercls{\aastexcls}
\documentclass[tighten, times, twocolumn]{aastex62}  

\if\papercls \apjcls
\usepackage{apjfonts}
\else\if\papercls \othercls
\usepackage{epsfig}
\usepackage{margin}
\usepackage{times}
\fi\fi
\usepackage{ifthen}
\usepackage{natbib}
\usepackage{amssymb, amsmath}
\usepackage{appendix}
\usepackage{etoolbox}
\usepackage[T1]{fontenc}
\usepackage{paralist}

\if\papercls \apjcls
\newcommand\aas{\ref@jnl{AAS Meeting Abstracts}}
\newcommand\dps{\ref@jnl{AAS/DPS Meeting Abstracts}}
\newcommand\maps{\ref@jnl{MAPS}}
\else\if\papercls \othercls
\usepackage{astjnlabbrev-jh}
\fi\fi

\if\papercls \aastexcls
\hypersetup{citecolor=blue, 
            linkcolor=blue, 
            menucolor=blue, 
            urlcolor=blue}  
\else
\usepackage[
bookmarks=true,           
bookmarksnumbered=true,   
colorlinks=true,          
citecolor=blue,           
linkcolor=blue,           
menucolor=blue,           
urlcolor=blue,            
linkbordercolor={0 0 1},  
pdfborder={0 0 1},
frenchlinks=true]{hyperref}
\fi

\if\papercls \othercls

\else

\fi

\providecommand{\adsurl}[1]{\href{#1}{ADS}}

\makeatletter
\patchcmd{\NAT@citex}
  {\@citea\NAT@hyper@{%
     \NAT@nmfmt{\NAT@nm}%
     \hyper@natlinkbreak{\NAT@aysep\NAT@spacechar}{\@citeb\@extra@b@citeb}%
     \NAT@date}}
  {\@citea\NAT@nmfmt{\NAT@nm}%
   \NAT@aysep\NAT@spacechar\NAT@hyper@{\NAT@date}}{}{}

\patchcmd{\NAT@citex}
  {\@citea\NAT@hyper@{%
     \NAT@nmfmt{\NAT@nm}%
     \hyper@natlinkbreak{\NAT@spacechar\NAT@@open\if*#1*\else#1\NAT@spacechar\fi}%
       {\@citeb\@extra@b@citeb}%
     \NAT@date}}
  {\@citea\NAT@nmfmt{\NAT@nm}%
   \NAT@spacechar\NAT@@open\if*#1*\else#1\NAT@spacechar\fi\NAT@hyper@{\NAT@date}}
  {}{}
\makeatother

\makeatletter
\DeclareRobustCommand{\lowcase}[1]{\@lowcase#1\@nil}
\def\@lowcase#1\@nil{\if\relax#1\relax\else\MakeLowercase{#1}\fi}
\makeatother

\DeclareSymbolFont{UPM}{U}{eur}{m}{n}
\DeclareMathSymbol{\umu}{0}{UPM}{"16}
\let\oldumu=\umu
\renewcommand\umu{\ifmmode\oldumu\else\math{\oldumu}\fi}
\newcommand\micro{\umu}
\if\papercls \othercls

\else

\fi

\let\oldsim=\sim
\renewcommand\sim{\ifmmode\oldsim\else\math{\oldsim}\fi}
\let\oldpm=\pm
\renewcommand\pm{\ifmmode\oldpm\else\math{\oldpm}\fi}
\newcommand\by{\ifmmode\times\else\math{\times}\fi}

\newbox{\wdbox}
\renewcommand\c{\setbox\wdbox=\hbox{,}\hspace{\wd\wdbox}}
\renewcommand\i{\setbox\wdbox=\hbox{i}\hspace{\wd\wdbox}}

\newcount\timect
\newcount\hourct
\newcount\minct
\newcommand\now{\timect=\time \divide\timect by 60
         \hourct=\timect \multiply\hourct by 60
         \minct=\time \advance\minct by -\hourct
         \number\timect:\ifnum \minct < 10 0\fi\number\minct}

\catcode`@=11

\newcommand\comment[1]{}

\newcommand\commenton{\catcode`\%=14}

\renewcommand\math[1]{$#1$}
\newcommand\mathshifton{\catcode`\$=3}

\let\atab=&
\newcommand\atabon{\catcode`\&=4}

\let\oldmsp=\sp
\let\oldmsb=\sb
\def\sp#1{\ifmmode
           \oldmsp{#1}%
         \else\strut\raise.85ex\hbox{\scriptsize #1}\fi}
\def\sb#1{\ifmmode
           \oldmsb{#1}%
         \else\strut\raise-.54ex\hbox{\scriptsize #1}\fi}
\newbox\@sp
\newbox\@sb
\def\sbp#1#2{\ifmmode%
           \oldmsb{#1}\oldmsp{#2}%
         \else
           \setbox\@sb=\hbox{\sb{#1}}%
           \setbox\@sp=\hbox{\sp{#2}}%
           \rlap{\copy\@sb}\copy\@sp
           \ifdim \wd\@sb >\wd\@sp
             \hskip -\wd\@sp \hskip \wd\@sb
           \fi
        \fi}
\def\msp#1{\ifmmode
           \oldmsp{#1}
         \else \math{\oldmsp{#1}}\fi}
\def\msb#1{\ifmmode
           \oldmsb{#1}
         \else \math{\oldmsb{#1}}\fi}

\def\supon{\catcode`\^=7}

\def\subon{\catcode`\_=8}

\def\supsubon{\supon \subon}

\newcommand\actcharon{\catcode`\~=13}

\newcommand\paramon{\catcode`\#=6}

\comment{And now to turn us totally on and off...}

\newcommand\reservedcharson{ \commenton  \mathshifton  \atabon  \supsubon 
                             \actcharon  \paramon}

\catcode`@=12
\reservedcharson

\if\papercls \apjcls

\else

\fi

\newcommand\tnm[1]{\tablenotemark{#1}}

\newcommand\chisq{\ifmmode{\chi\sp{2}}\else\math{\chi\sp{2}}\fi}
\newcommand\redchisq{\ifmmode{ \chi\sp{2}\sb{\rm red}}
                    \else\math{\chi\sp{2}\sb{\rm red}}\fi}
\newcommand\Teq{\ifmmode{T\sb{\rm eq}}\else$T$\sb{eq}\fi}
\newcommand\mjup{\ifmmode{M\sb{\rm Jup}}\else$M$\sb{Jup}\fi}
\newcommand\rjup{\ifmmode{R\sb{\rm Jup}}\else$R$\sb{Jup}\fi}
\newcommand\msun{\ifmmode{M\sb{\odot}}\else$M\sb{\odot}$\fi}
\newcommand\rsun{\ifmmode{R\sb{\odot}}\else$R\sb{\odot}$\fi}
\newcommand\mearth{\ifmmode{M\sb{\oplus}}\else$M\sb{\oplus}$\fi}
\newcommand\rearth{\ifmmode{R\sb{\oplus}}\else$R\sb{\oplus}$\fi}

\shorttitle{Goel {\em et al.}}
\shortauthors{Goel {\em et al.}}

\usepackage{xcolor}
\usepackage{newtxtext,newtxmath}

\begin{document}

\newcommand*{\red}{\textcolor{red}}
\newcommand*{\blue}{\textcolor{blue}}

\title{A Comprehensive Analysis of WISE Mid-Infrared Colors for Obscured AGN Selection}

\author{Anika Goel}
\affiliation{Indiana University, 727 East 3rd St. Swain West 318, Bloomington, IN 47405-7105, USA}

\author{Samir Salim}
\affiliation{Indiana University, 727 East 3rd St. Swain West 318, Bloomington, IN 47405-7105, USA}

\author{Sara L.\ Ellison}
\affiliation{University of Victoria, 3800 Finnerty Road, Victoria BC  V8P 5C2, Canada}

\author{Shobita Satyapal}
\affiliation{George Mason University, 400 University Drive, Fairfax, VA 22030, USA}

\author{Sheyda Salehirad}
\affiliation{Montana State University, Bozeman, MT 59717, USA}

\author{Robert W.\ Bickley}
\affiliation{University of Victoria, 3800 Finnerty Road, Victoria BC  V8P 5C2, Canada}

\author{Christopher J.\ Agostino}
\affiliation{Indiana University, 727 East 3rd St. Swain West 318, Bloomington, IN 47405-7105, USA}


\email{anikgoel@iu.edu}


\begin{abstract}
In this paper, we investigate the robustness of WISE mid-IR color selection ($W1-W2$) for identifying obscured (Type 2) active galactic nuclei (AGNs) at low redshift ($z<0.3$), using a sample of $\sim$360,000 SDSS galaxies classified via emission lines into Seyfert 2 (Sy2), LINER, and star-forming (BPT-SF) galaxies. 
We find that the K-correction is essential to remove non-AGN contamination, and once applied the simple $W1-W2>0.5$ selection emerges as optimal in terms of purity and completeness of AGN selection. However, we confirm that even this lenient cut selects only $\sim 13\%$ of Sy2 galaxies and that achieving $W1-W2>0.5$ requires AGN contributing $\geq 75\%$ of the total infrared luminosity, which is uncommon. Although mid-IR-selected Sy2s tend to be luminous, the high [OIII] luminosity does not guarantee red $W1-W2$ (nor does any other tested global or NLR-scale parameter), suggesting the critical role of obscuration on smaller scales.
$<1\%$ of BPT-SF systems (but making $\approx 20\%$ of all mid-IR selected galaxies) exhibit $W1-W2>0.5$ colors. Such colors cannot be reproduced by models of star-heated dust alone. Red BPT-SFs tend to have higher $W4$ luminosities than expected from SF, indicating true AGNs. Intriguingly, mid-IR AGNs in massive bulges ($M_{\mathrm{bulge}} \gtrsim 10^{10} M_{\odot}$) predominantly (84\%) manifest themselves as BPT-AGNs, whereas those in low-mass bulges ($\lesssim 10^{10} M_{\odot}$)  mostly (60\%) manifest as BPT-SF. This BPT-AGN vs.\ BPT-SF dichotomy does not extend to total stellar mass. We conclude that although the mid-IR AGN selection is incomplete, its strength lies in identifying optically inconspicuous AGNs with low-mass bulges, regardless of the total mass. 

\end{abstract}


\section{Introduction}
\label{sec:intro}

Most massive galaxies in the universe are known to host a supermassive black hole (SMBH) at their centers \citep{Kormendy1995, Kormendy2013ARA&A..51..511K}. When these SMBHs are actively accreting gaseous material, they can be observed as Active Galactic Nuclei (AGN). AGNs manifest in a range of types, which can be broadly explained by the differences in orientation, accretion rates, presence or absence of jets and the environment of host galaxies \citep{Padovani_2017}. The unified model of AGN is the leading theory for interpreting this diversity. According to this framework, the observed differences in AGN population arise due to orientation effects --- i.e., how the central engine and its surrounding structures are viewed along our line of sight \citep{Antonucci1993ARA&A..31..473A, Urry1995PASP..107..803U}. In this picture, an AGN is classified as unobscured, or ``Type 1'', when the emission from the accretion disk is directly visible, and as obscured, or ``Type 2'', when the emission from the accretion disk is blocked by the surrounding dust torus \citep{krolik1988ApJ...329..702K, nenkova2008ApJ...685..160N}. Orientation also explains the presence or absence of the Broad Line Region (BLR) in optical spectra, which arises from fast moving gas clouds. The Narrow Line Region (NLR), on the other hand, is usually visible for both Type 1 and Type 2 AGNs. This distinction arises from the relative distances of the BLR and the NLR with respect to the central SMBH: the BLR lies closer to the gravitational influence of the SMBH and shows a wider velocity distribution ($\sim 1000-10,000 \ \mathrm{km\,s^{-1}}$), whereas the NLR lies further out, with the typical velocity distributions of $\sim 100-500\ \mathrm{km\, s^{-1}}$ \citep{hickox2018ARA&A..56..625H, Padovani_2017}.

Understanding the role of SMBH growth in galaxy evolution requires a complete census of AGN hosts spanning the full diversity of galaxy types. Such a census is particularly lacking for lower-mass galaxies ($M_*\lesssim10^{10} M_{\odot}$), where AGN activity and the presence of central black holes remains poorly constrained \citep{Greene2020ARA&A..58..257G, shobita2014ApJ...784..113S, Sturm2025arXiv250109791S}. Even for massive hosts, building a complete, unbiased sample of AGNs continues to be a major challenge for current observational studies, especially for obscured AGNs, which might be harder to detect.

Optical emission-line diagnostics, and most famously the Baldwin-Phillips-Terlevich (BPT) diagram, provide one of the most powerful spectroscopic tools for identifying obscured AGN \citep{BPT1981PASP...93....5B,Kauffmann2003, Kewley2006MNRAS.372..961K}. Because the UV to near-IR region is rich in strong forbidden and permitted lines, whose ratios tell us about the ionization state and excitation conditions of the gas, these diagrams are able to separate galaxies or regions where the ionizing source is predominantly an AGN or young stars (HII regions). Emission line diagnostics are especially valuable for identifying obscured AGNs, where the dominant source of nebular ionization is an accretion disk enshrouded by NLR clouds. However, the effectiveness of BPT line diagnostics can be affected by the following factors:
\begin{enumerate}
    \item Dust can attenuate or extinguish the emission lines produced by AGNs (NLR emission), and strong star formation might additionally make them harder to detect. Alternatively, emission lines produced by AGNs might be intrinsically weak even in otherwise powerful AGNs \citep{agostino2023ApJ...943..174A, fiore2000NewA....5..143F, M2000Natur.404..459M, barger2001AJ....121..662B, szo2004ApJS..155..271S, comastri2002ApJ...571..771C, brusa2003A&A...409...65B, civano2007A&A...476.1223C, rigby2006ApJ...645..115R, trump2009ApJ...706..797T, trouille2010ApJ...722..212T, koss2017ApJ...850...74K}. 
    
    \item AGN identification in dwarf/low-mass galaxies is particularly challenging using the BPT diagram \citep{cann2019ApJ...870L...2C}, and indeed a large fraction of galaxies that show signatures of AGN from the coronal lines, He II or optical or mid-IR variability are found in the star-forming (non-AGN) branch of the BPT diagram (e.g., \citealt{baldassare2020ApJ...896...10B, salehirad2022ApJ...937....7S, 2022ApJ...936..104W, wasleske2024ApJ...971...68W}).
    
    \item BPT line diagnostics separate both Seyfet 2s and LINER galaxies from those that are star-forming. However, the true nature of ionization in LINER galaxies is not fully understood. Early studies characterized LINER galaxies as AGN \citep{heckman1980A&A....87..152H, Ho1993ApJ...417...63H} motivated by the presence of strong X-ray emission and broad emission lines. However, other studies proposed a post-AGB stellar origin of ionization (\citealt{1994A&A...292...13B, 2008MNRAS.391L..29S, 2010MNRAS.402.2187S, 2011MNRAS.413.1687C, 2012ApJ...747...61Y, 2013A&A...558A..43S, 2016MNRAS.461.3111B, 2016A&A...588A..68G, 2017A&A...599A.141J, 2017MNRAS.466.3217Z, 2019AJ....158....2B}) or shocks (\citealt{1995ApJ...455..468D, 1996ApJS..102..161D, 2001ApJ...556..121K, 2010ApJ...721..505R, 2011ApJ...734...87R, 2014ApJ...781L..12R, 2017MNRAS.470.4974D, 2018ApJ...864...90M, 2019MNRAS.485L..38D}). 
\end{enumerate}

While the BPT diagnostic is generally very reliable, the necessity for spectroscopy and the potential limitations listed above have motivated the development of complementary approaches to identify AGNs, and especially obscured ones \citep{hickox2018ARA&A..56..625H}. Since the dust obscuring the AGN absorbs high-energy radiation from the accretion disk and re-emits it in the mid-infrared (mid-IR, 3- 30 $\micro$m), it provides an alternative avenue for identifying obscured AGNs. Mid-IR color diagnostics, especially those using the photometry from Wide-Field Infrared Survey Explorer (WISE; \citealt{wright2010AJ....140.1868W}), are widely used to identify AGNs based on differences in their mid-IR SEDs (e.g., \citealt{Stern2012ApJ...753...30S, Assef2013ApJ...772...26A, Assef2018ApJS..234...23A, Mateos2012MNRAS.426.3271M, Jarrett2011ApJ...735..112J, Yan2013AJ....145...55Y}). 

\citet{Stern2012ApJ...753...30S} proposed a color cut using $W1 - W2 \geq 0.8$ (where W1 and W2 are the 3.4 $\micro$m and 4.6 $\micro$m WISE bands) to select AGNs out to a redshift of $z \sim 3$, with high reliability when compared to the Spitzer mid-IR AGN selection.
\citet{Assef2013ApJ...772...26A} extended this work by refining the color criteria for fainter sources, improving AGN selection in deeper datasets.  They also found the $W1-W2>0.5$ selection to be 90\% complete with respect to the Spitzer IRAC-selected AGN candidates. This more encompassing selection was used by \citet{shobita2014ApJ...784..113S}, \citet{Blecha2018MNRAS.478.3056B} and many subsequent works, and we will be focusing the most on it in this work.

\citet{Assef2013ApJ...772...26A} also found that their $W1-W2$ criterion is sensitive to galaxies with strong AGN contributions ($\hat{a}>0.5$), where $\hat{a}$ is parameterized as the ratio of the AGN luminosity compared to the total host galaxy luminosity and AGN luminosity. As a result, the mid-IR selection is known to miss a fair fraction of X-ray selected AGN, especially of lower X-ray luminosities. This has been highlighted already in Spitzer studies \citep{Barmby2006ApJ...642..126B, Donley2007ApJ...660..167D}, and was confirmed for WISE selection \citep{Mateos2012MNRAS.426.3271M,lamassa2019ApJ...876...50L}. There are indications that the mid-IR AGN selection might be even less complete with respect to the optically-selected obscured AGN, i.e., the emission-line (BPT) AGNs \citep{shobita2014ApJ...784..113S,sartori2015MNRAS.454.3722S,Hviding2022AJ....163..224H, Bickley2024MNRAS.533.3068B}, and that in this case too the completeness may depend on AGN luminosity \citep{Ellison2025OJAp....8E..12E}. In the current work we aim to explore this relationship between mid-IR and emission-line AGN selection more quantitatively, focusing on large samples and at lower redshifts ($z<0.3$). 

Previous studies using Spitzer have found that mid-IR AGN selection is generally free from non-AGN galaxy contamination, especially at $z<1$ \citep{2012ApJ...748..142D}. However, even a small number of apparent non-AGN ``contaminants'' could be important for understanding the processes that heat the dust \citep{Izotov2011A&A...536L...7I}. Thus, some studies have questioned the reliability of the mid-IR color selection as a method to identify AGNs. \citet{shobita2014ApJ...784..113S} found that some bulgeless galaxies have red WISE colors indicative of an AGN, but not showing the line ratios characteristic of AGN. \cite{Connor2016MNRAS.463..811O} studied the WISE colors of $\sim$ 315,000 nearby ($z<0.1$) galaxies and found that 86\% of lower-mass galaxies ($M < 10^{9} \ M_{\odot}$) with $W1-W2>0.5$ were not classified by BPT as AGNs, but rather as star forming. \citet{lamassa2019ApJ...876...50L} found 50\% of $z<0.5$ WISE AGNs (selected with the \citet{Assef2018ApJS..234...23A} 75\% reliability criteria)to be classified as SF in the BPT diagram. 

Although there is no reason why the overlap between the different AGN selection criteria should be perfect, such discrepancies do bring into question the reliability of any given method, and still require some explanation. Several potential reasons for the discrepant AGN status between mid-IR and optical-line selection have been suggested in literature. \cite{Hainline2016ApJ...832..119H} suggested that galaxies with low metallicities might give rise to AGN-like mid-IR color in low-mass hosts. 
Another factor potentially affecting the mid-IR AGN selection is the presence of starburst. So in both of these scenarios, these mid-IR selected objects would only mimic AGN signatures, but not be real AGNs. However, \cite{Blecha2018MNRAS.478.3056B} performed a theoretical study, finding that merger-induced nuclear starburst can mimic AGN $W1-W2$ color, but only briefly. Also based on modeling, \cite{Shobita2018ApJ...858...38S} showed that extreme starburst galaxies can in theory produce $W1-W2$ colors similar to those of AGNs, due to the intense dust heating, but that this requires unrealistically high ionization parameters. A follow-up of 11 mid-IR selected AGN candidates from \citet{Hainline2016ApJ...832..119H}, showed no X-ray AGN signatures in all five BPT SF galaxies, whereas the BPT AGNs were confirmed as X-ray AGNs \citep{latimer2021ApJ...914..133L}. More recently, \cite{Sturm2025arXiv250109791S} examined with the HST four mid-IR identified AGN candidates in dwarf galaxies, also first identified by \citet{Hainline2016ApJ...832..119H}, and found that these systems host rare, young, and massive nuclear star clusters that might be responsible for extreme dust heating. However, \citet{Doan2025ApJ...987...99D} observed with HST and JWST another extremely metal-poor dwarf galaxy with discrepant mid-IR and BPT AGN status, and although they also find a young nuclear star cluster, they note that an unresolved ($<5$ pc) source, possibly an AGN, is responsible for the mid-IR continuum. In some other cases, galaxies with red $W1-W2$ color are more clearly confirmed as AGNs using some other method, despite their non-AGN emission line classification \citep{shobita2016ApJ...827...58S,secrest2015ApJ...798...38S}, especially in cases where their $W2-W3$ color is not particularly red \citep{Shobita2018ApJ...858...38S, Harish2023ApJ...945..157H}. Notably, 17 out of 20 dwarf/low-mass galaxies ($\log M_*<9.6$) that show mid-IR variability (and typically have red $W1-W2$) are actually found in the star-forming region of the BPT diagram \citep{Aravindan2024ApJ...975...60A}.   

The above literature review highlights the complexity of multi-wavelength AGN selection.  In particular, whilst mid-IR color selection is a potentially powerful tool, there are concerns for both completeness and contamination.  This leaves us with many open questions. How often do the simple WISE cuts fail to recover obscured or low-luminosity AGN? How reliably do mid-IR cuts identify obscured AGN systems? Are red mid-IR sources AGNs even when their emission-line classification does not indicate an AGN? These open questions highlight the need for a deeper evaluation of mid-IR selection methods, particularly using low-$z$, large-survey datasets. A more nuanced understanding of the strengths and limitations of mid-IR techniques is essential to constructing an unbiased AGN census and to interpreting the demographics of SMBH growth across cosmic time.

The goal of this paper is to evaluate type 2 AGN selection with WISE mid-IR color on a large and diverse galaxy sample from the GALEX-SDSS-WISE Legacy Catalog (GSWLC), a catalog of SED fitting derived parameters of $z<0.3$ galaxies. GSWLC is particularly well suited for this analysis as it provides accurate star formation rates and stellar masses, and is associated with high-quality optical emission-line measurements from SDSS, which allow robust emission-line classification. Examining the WISE AGN selection with this dataset provides an opportunity to evaluate its reliability and completeness in a broader context. 

\begin{deluxetable*}{p{2cm} p{1cm} r r r r p{6cm}}[htbp]  
\tabletypesize{\footnotesize}
\tablecaption{Sample Selection and Emission-line Classification \label{table:sample B}}

\tablehead{
    \colhead{\raggedright Category} & 
    \colhead{\raggedright Selection\tnm{a}} & 
    \colhead{\raggedright $N\tnm{b}$} & 
    \colhead{\raggedright $N$(WISE)\tnm{c}} & 
    \colhead{\raggedright Fraction with} & 
    \colhead{\raggedright Contribution to} & 
    \colhead{\raggedright Category description} \\
    \colhead{} & 
    \colhead{} & 
    \colhead{} & 
    \colhead{} & 
    \colhead{\raggedright $(W1-W2)_{z=0}$} & 
    \colhead{\raggedright $(W1-W2)_{z=0}$} & 
    \colhead{}    \\
    \colhead{} & 
    \colhead{} & 
    \colhead{} & 
    \colhead{} & 
    \colhead{\raggedright $> 0.5$} & 
    \colhead{\raggedright $> 0.5$} & 
    \colhead{}   
}

\startdata
SF (= BPT-SF)  & C2=1  & 90,421   & 86,096   & 0.62\%  & 20.74\% & BPT star-forming (all four BPT lines have SNR>2)
\\
Additional SF  & C2=2  & 10,955   & 10,784   & 0.31\%  & 1.27\% & Classified as SF based on log ([NII]/H$\alpha) < -0.4$\, (H$\beta$ or [OIII] too weak (SNR<2) for full BPT classification) 
\\
Seyfert 2  & C2=5 AND C3=1  & 7,055    & 7,032    & 13.30\%  & 36.04\% & BPT-AGN sub-classified as a Seyfert 2 based on seven lines/doublets (all with SNR>2) with SF contribution removed 
\\
LINER  & C2=5 AND (C3=2 OR C3=3)  & 38,203   & 38,102   & 1.02\%  & 15.00\% & BPT-AGN sub-classified as a LINER based on seven lines/doublets (all with SNR>2) with SF contribution removed
\\
Additional LINER or Sy2  & C2=3 OR (C2=5 AND C3=9)  & 101,136  & 100,903  & 0.47\%  & 18.35\% & BPT-AGN lacking Sy2/LINER sub-classification, or selected using log ([NII]/H$\alpha)  > -0.35$ when H$\beta$ or [OIII] are too weak (SNR<2) for full BPT classification
\\
Weak or No Lines  & C2=2  & 76,425   & 75,874   & 0.18\%  & 5.13\% & Could not be classified (either [NII] or H$\alpha$ too weak (SNR<2))
\\
Uncertain  & C2=4  & 15,610 & 15,546  & 0.58\%  & 3.47\% & Galaxies in between SF/AGN demarcation lines ($-0.4<$ log ([NII]/H$\alpha)  < -0.35$, with both lines SNR>2)
\\
TOTAL  & - & 324,195  & 318,791  & 0.81\%  & 100\% & Complete sample \\
\hline
\enddata

\tablecomments{Table note a: C2 and C3 refer to columns 2 and 3 in Table 1 in \citet{Agostino_2021}. Table note b: Number of objects in a given category. Table note c: Objects with valid W1 and W2 photometry and W1-W2 error < 0.1.}
\end{deluxetable*}

\section{Sample and Data}
\label{sec:observations}
\subsection{Data}
The principal physical properties (including stellar mass, SFR and dust attenuation) for galaxies in our sample are drawn from the GALEX-SDSS-WISE Legacy Catalogs 1 and 2 (hereafter GSWLC-1 and GSWLC-2). GSWLC-1 \citep{gswlc-1} includes galaxy properties of \sim 700,000 optically selected galaxies at $z<0.3$, obtained by SED fitting of UV/optical photometry using CIGALE \citep{cigale}. GSWLC-1 separately lists the SFRs from WISE 22 $\micro$m photometry (when available), derived from luminosity-dependent IR dust SED templates. 

GSWLC-2 \citep{2018ApJ...859...11S} contains identical sources and uses the same photometry as GSWLC-1, however, the IR luminosity derived from 22 $\micro$m WISE (or 12 $\micro$m, when 22 $\micro$m is unavailable) is used within the SED fitting process, concurrently with UV/optical photometry. Although GSWLC-2 SFRs are generally more precise than GSWLC-1 ones, any uncorrected AGN contribution to 22 (or 12) $\micro$m flux might bias them, which is why in most cases we use SFRs from GSWLC-1.  

We use mid-IR photometry from WISE, a space-based mission which observed the full sky in four different photometric bands, at 3.4 ($W1$), 4.6 ($W2$), 12 ($W3$) and 22 ($W4$) $\micro$m. Specifically, this work uses the unWISE catalog\footnote{\texttt{https://catalog.unwise.me/}} which performed flux measurements based on SDSS DR10 detection positions and profile shapes \citep{lang14}. We convert unWISE fluxes to Vega magnitudes.

We use line flux measurements and velocity dispersions from the MPA/JHU catalog derived following \cite{Tremonti_2004} and obtained for galaxies in SDSS DR7 \citep{Abazajian2009ApJS..182..543A}. We corrected all emission-line fluxes for internal dust attenuation using the Balmer decrement. Assuming case B recombination (intrinsic $L_{\mathrm{H}\alpha}/L_{\mathrm{H}\beta}=2.86$) and the \cite{Cardelli1989ApJ...345..245C} extinction curve, the $V$-band extinction towards nebular gas is given as

\begin{equation}
A_{V,\mathrm{neb}} = 7.23 \, \log_{10} \left( \frac{L_{\mathrm{H}\alpha}/L_{\mathrm{H}\beta}} {2.86} \right).
\end{equation}

\noindent where the factor 7.23 arises from the $2.5/[k(\mathrm{H\beta})- k(\mathrm{H\alpha})]$ term. The corrected fluxes are then,

\begin{equation}
F_{\mathrm{corr}}(\lambda) = F_{\mathrm{obs}}(\lambda) \times 10^{0.4 \, A_{V,\mathrm{neb}} \, k(\lambda)},
\end{equation}

\noindent where \( k(\lambda) \) corresponds to the reddening coefficients: \( k(5007\AA) = 1.120 \) for [OIII]~$\lambda5007$, \( k(4861\AA) = 1.164 \) for H$\beta$, \( k(6584\AA) = 0.815 \) for [NII]~$\lambda6584$ and \( k(6563\AA) = 0.818 \) for H$\alpha$. 

The velocity dispersions we use are from the widths of the forbidden lines. The results are qualitatively the same if we were to use stellar velocity dispersions, but those measurements tend to be less precise. 

We compute the bulge stellar mass according to the relation:
\[
\log \ M_{\mathrm{bulge}} = \log \ M_* + \log \ (B/T)
\]
where $M_*$ is the total stellar mass from GSWLC-M2 and $B/T$ is the bulge-to-total stellar mass ratio obtained from Sersic plus exponential profile bulge/disk decompositions of \citet{mendel14}. To avoid unphysical ratios, we take $T$ to be the sum of the bulge and disk masses, both from \citet{mendel14}. We also point out that because the decomposition always includes two components, the bulgeless galaxies are not identified as a separate category, but will have some, presumably small, $B/T$ ratio.

The X-ray data come from the tenth data release of the fourth XMM-Newton source catalog (4XMM-Newton DR10; \citealt{2020A&A...641A.136W}), which includes sources from 11647 XMM-Newton EPIC observations. These were matched to SDSS and X-ray luminosities were derived corresponding to 0.5--10 keV band. Details are provided in \citet{Agostino_2023ApJ...943..174A}.

\subsection{Sample Selection and Emission-line Classification}
\label{sample}

GSWLC, which forms the basis of our sample, consists of three catalogs based on the depth of the UV photometry: shallow (GSWLC-A), medium (GSWLC-M) and deep (GSWLC-D) covering 88\%, 49\% and 7\%  of the SDSS spectroscopic sample, respectively. For this project, we only utilize the medium-deep catalog (GSWLC-M), due to its larger sample size compared to GSWLC-D, and its more precise SFR estimates compared to GSWLC-A. From GSWLC-2 we exclude type 1 (broad-line) AGNs (mostly Seyfert 1s), based on the `QSO' classification in the SDSS spectroscopic pipeline, leaving $\sim$324,000 galaxies.

\begin{figure}[h!]
\centering
\includegraphics[width=0.4\textwidth, trim={10 30 10 30}]{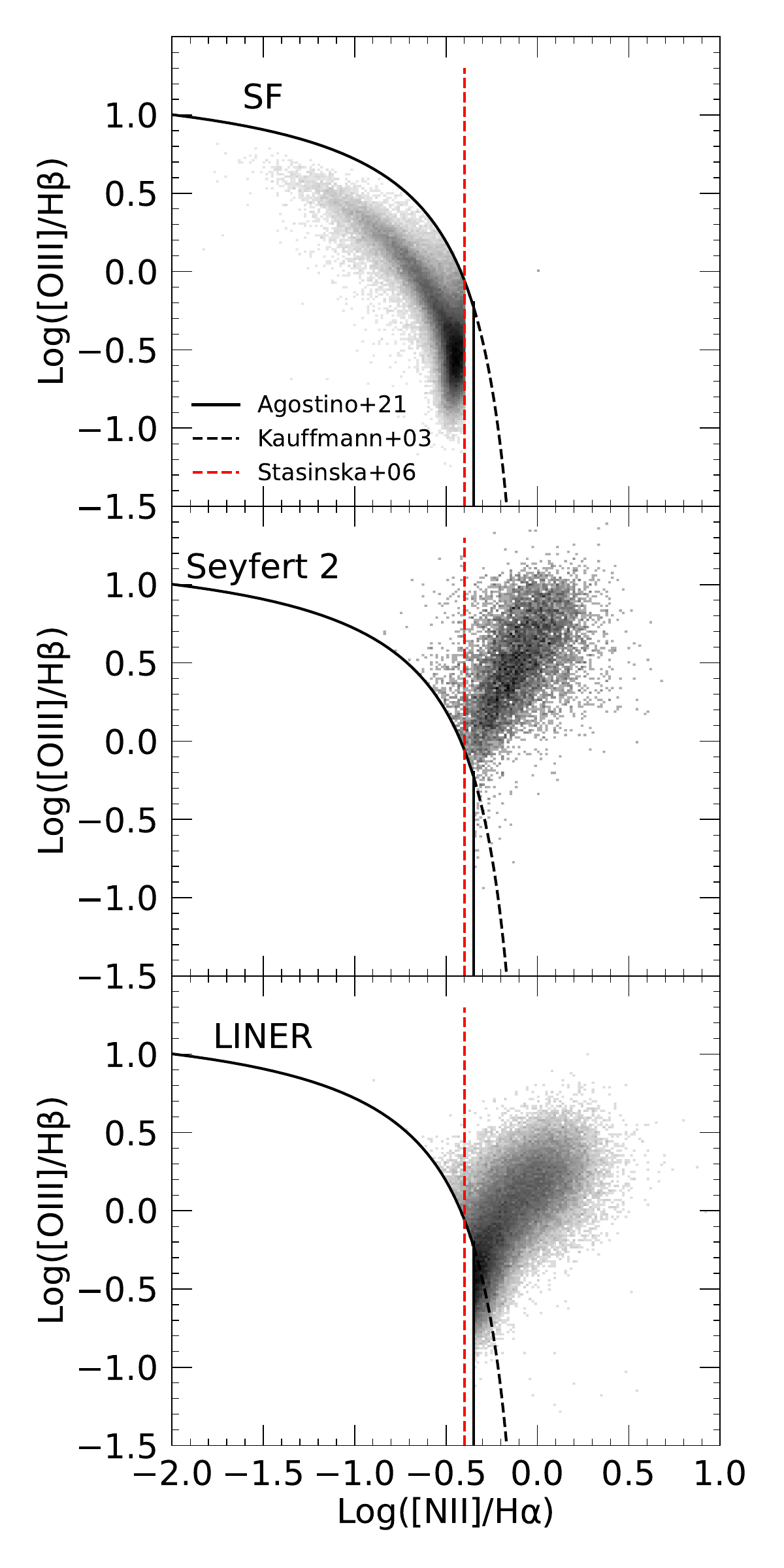} 
\caption{BPT diagrams for star-forming (SF), Seyfert 2 (Sy2), and LINER galaxies from the GSWLC-M2 sample, classified following \citet{Agostino_2021}. Galaxies are separated into SF and AGN (Sy2/LINER) using the modified demarcation line that combines \citet{Kauffmann2003} demarcation (black curve) and the \citet{Stasinska_2006} vertical cut (red dashed line). SNR > 2 is required in the four diagnostic lines. AGNs are subdivided into Sy2 and LINER galaxies using additional lines and the $k$-means scheme of \citet{Agostino_2021}. Galaxies in the narrow vertical slice between the red \citet{Stasinska_2006} vertical cut and black vertical \citet{Agostino_2021} line are discarded and marked ``uncertain''.}
\label{fig:bpt}
\end{figure}

To investigate how WISE color selection behaves across different optical emission-line galaxy types, we adopt the classification framework developed by \citet{Agostino_2021}. As shown in Figure \ref{fig:bpt}, this method applies an AGN demarcation line that combines the vertical cut (log([NII]/H$\alpha=-0.35$) with the \cite{Kauffmann2003} curved demarcation line. The new demarcation shifts a portion of galaxies classified as SF under the original \cite{Kauffmann2003} line into the AGN (Seyfert/LINER) category, resulting in a cleaner division between the two populations. Furthermore, \cite{Agostino_2021} determined that the \citet{Stasinska_2006} vertical cut of log([NII]/H$\alpha) = -0.4$ served as the optimal boundary to select SF (non-AGN) galaxies in the sample. The objects between the boundaries of $-0.4$ and $-0.35$ were discarded, and marked as ``Uncertain''.

We require the signal-to-noise ratio (SNR) to be greater than 2 in all four diagnostic lines to select SF and AGN categories. This SNR cut results in well-defined branches on the BPT diagram. For galaxies with weaker [OIII] or H$\beta$ (SNR < 2), but [NII] and H$\alpha$ with SNR $>2$, the [NII]/H$\alpha$ is alone used to separate the populations into SF and AGN using.  Galaxies with log([NII]/H$\alpha) < -0.4$ are labeled ``Additional SF'', while galaxies with log([NII]/H$\alpha) > -0.35$ are labeled ``Additional LINER or Sy2''.

The AGN population is further classified in \cite{Agostino_2021} into Seyfert 2s (Sy2), Hard-LINERs (H-LINER) and Soft-LINER (S-LINER), using the $k$-means multi-dimensional clustering, based on the four BPT lines plus additional emission lines: [SII] and [OII] doublets, as well as [OI], when detected. The clustering is performed in a seven-dimensional space defined by the emission lines after subtracting the star-forming contribution from HII regions. In this method, Sy2 galaxies occupy the high-ionization, harder radiation end of the AGN branch, while LINER galaxies are characterized by the lower-ionization branch. This method effectively separates AGN subtypes based on their emission-line properties and provides a cleaner separation of AGN subtypes than the traditional diagnostic diagrams. In this work we will not differentiate between the two types of LINERs.

Finally, there is a population of galaxies which could not be classified because even H$\alpha$ or [NII] fall below the SNR threshold, and are labeled ``Weak or No Lines''.

To summarize, the parent sample consists of \sim  324,000 galaxies, which are categorized into:\sim 7000 Sy2s, \sim 38,000 LINERs, \sim 86,000 SF galaxies, \sim 101,000 ``Additional LINER or Sy2'', $\sim$11,000 ``Additional SF'',  and \sim 76,000 unclassifiable galaxies with weak or no emission lines (``Weak or No Lines''). The exact values are given in Table \ref{table:sample B}. For the majority of our discussions we will focus on Sy2, LINERs and SF galaxies only. These categories are shown in Figure \ref{fig:bpt}. It should be emphasized that by calling the category that does not show the signatures of an AGN (from the BPT perspective) star-forming (SF), we just follow the usual nomenclature. This is not meant to mean that the galaxies on the AGN branch (Sy2s and LINERs) do not have SF present in them.  
  
To arrive at our final sample, we require valid W1 and W2 photometry, as well as an error on $W1-W2$ color to be $< 0.1$. After this cut we are left with 95.4\% of the original sample, as shown in Table \ref{table:sample B}. This is our principal sample.

For part of the analysis we also use X-ray AGNs, identified by \cite{Agostino_2023ApJ...943..174A} using the methodology of \cite{Agostino_2019ApJ...876...12A}, which defines X-ray AGNs as galaxies exhibiting an X-ray luminosity of more than 0.6 dex above that predicted from the SFR-$L_{X,0.5-10 KeV}$ relation of \cite{Ranali_2003A&A...399...39R}. Out of 712 X-ray sources matched to the GSWLC-M2 catalog, 638 were selected as AGN candidates, spanning $L_X = 10^{40} \ \text{to} \ 10^{45}$ erg s$^{-1}$, with roughly half below $L_X= 10^{42}$ erg s$^{-1}$, a threshold commonly used to define robust AGNs \citep{2005MNRAS.358..131G}. To ensure that the sample consists of genuine AGN, 126 sources with extended X-ray emission were removed. These sources likely trace diffuse hot gas and can mimic AGN signatures if extended. This selection methodology leaves 474 X-ray AGNs. Finally, requiring valid $W1$ and $W2$ photometry with $W1-W2$ errors $< 0.1$ removed one additional object, leaving a final sample of 473 X-ray AGNs. This number is much smaller than the number of optically selected AGNs primarily because the parent XMM catalog covers a relatively small portion of SDSS, and thus the GSWLC.

\section{Results}
\label{sec:result}

\subsection{K-correction of WISE colors}
\label{subsec:res_k_correct}

\begin{figure}[htb!]
\centering
\includegraphics[width=0.436\textwidth, trim={40 30 40 15}]{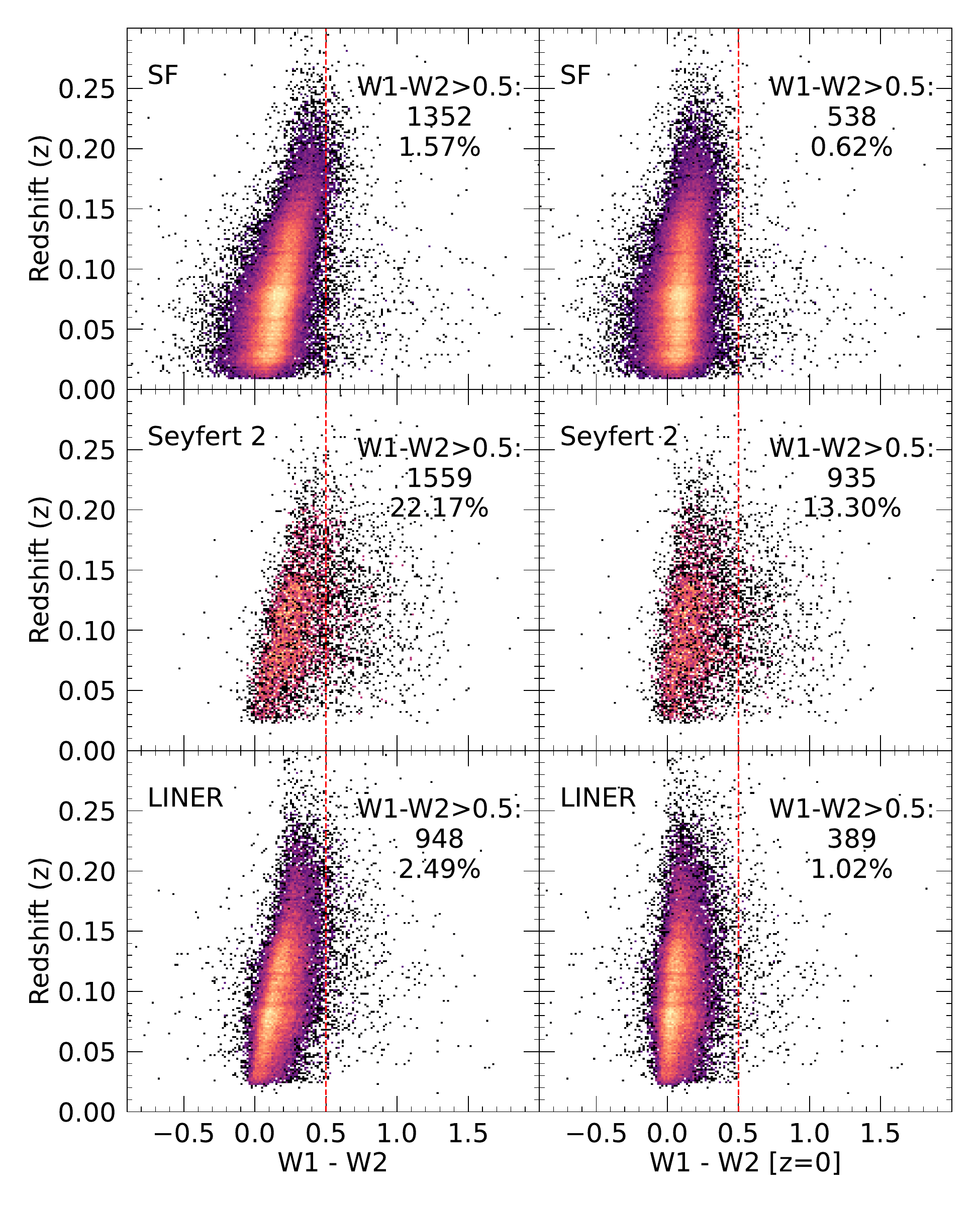} 
\caption{Redshift versus $W1-W2$ color for SF (top), Sy2 (middle), and LINER (bottom) galaxies. Left panels show the observed colors, whereas the right panels show rest-frame $W1-W2$ after the K-correction. Vertical line marks the AGN selection cut at $W1-W2=0.5$. A redshift-dependent tilt in left panels shifts higher-redshift objects toward redder colors and contributes to the contamination above the cut. This contamination is largely eliminated after the K-correction (right panels).}
\label{fig:redshift}
\end{figure}

The AGN selection threshold at $W1-W2=0.5$, proposed by \cite{Assef2013ApJ...772...26A}, is usually applied on {\it observed} magnitudes, rather than the rest-frame ones. Here we wish to test if not applying the K-correction is justified. The left panels of Figure \ref{fig:redshift} show the redshift vs.\ $W1-W2$ color for SF galaxies, Sy2s, and LINERs. The vertical line indicates the AGN selection threshold at $W1 - W2 = 0.5$. From here on, we will refer to galaxies with WISE colors $W1-W2 >0.5$ as ``red'' galaxies, and galaxies with colors $W1-W2<0.5$ as ``blue'' galaxies.  Each panel also reports the number and fraction of galaxies (in a given category) above this threshold.

In all three panels there is a systematic reddening of $W1-W2$ with increasing redshift. This trend suggests that the observed $W1-W2$ color is not constant with redshift, but instead drifts upward due to the K-correction effects, causing higher-redshift galaxies to appear artificially redder in color. This is most clearly seen in the SF panel, where a significant fraction of galaxies crosses the  $W1-W2=0.5$ line at $z>0.2$, indicating that some star-forming (i.e., non-AGN according to the BPT) galaxies would be incorrectly selected as mid-IR AGN. Although the overall fraction of BPT SF galaxies that would have a discordant classification as mid-IR AGN is not high (1.6\%), this might still lead to a large absolute number of false mid-IR AGN. 

To mitigate this bias and ensure that the color cut remains consistent across the redshifts, we K-correct the $W1 - W2$ color as follows:
\vspace{-0.5em}

\begin{equation} \label{eq:kcorr}
(W1 - W2)_{[z=0]} =
\begin{cases}
(W1 - W2) - 1.0\,z, & \text{if } z \le 0.2 \\
(W1 - W2) - 0.2, & \text{if } 0.2 < z < 0.3.
\end{cases}
\end{equation}

\begin{figure}[h!]
\centering
\includegraphics[width=0.45\textwidth, trim={45 20 30 20}]{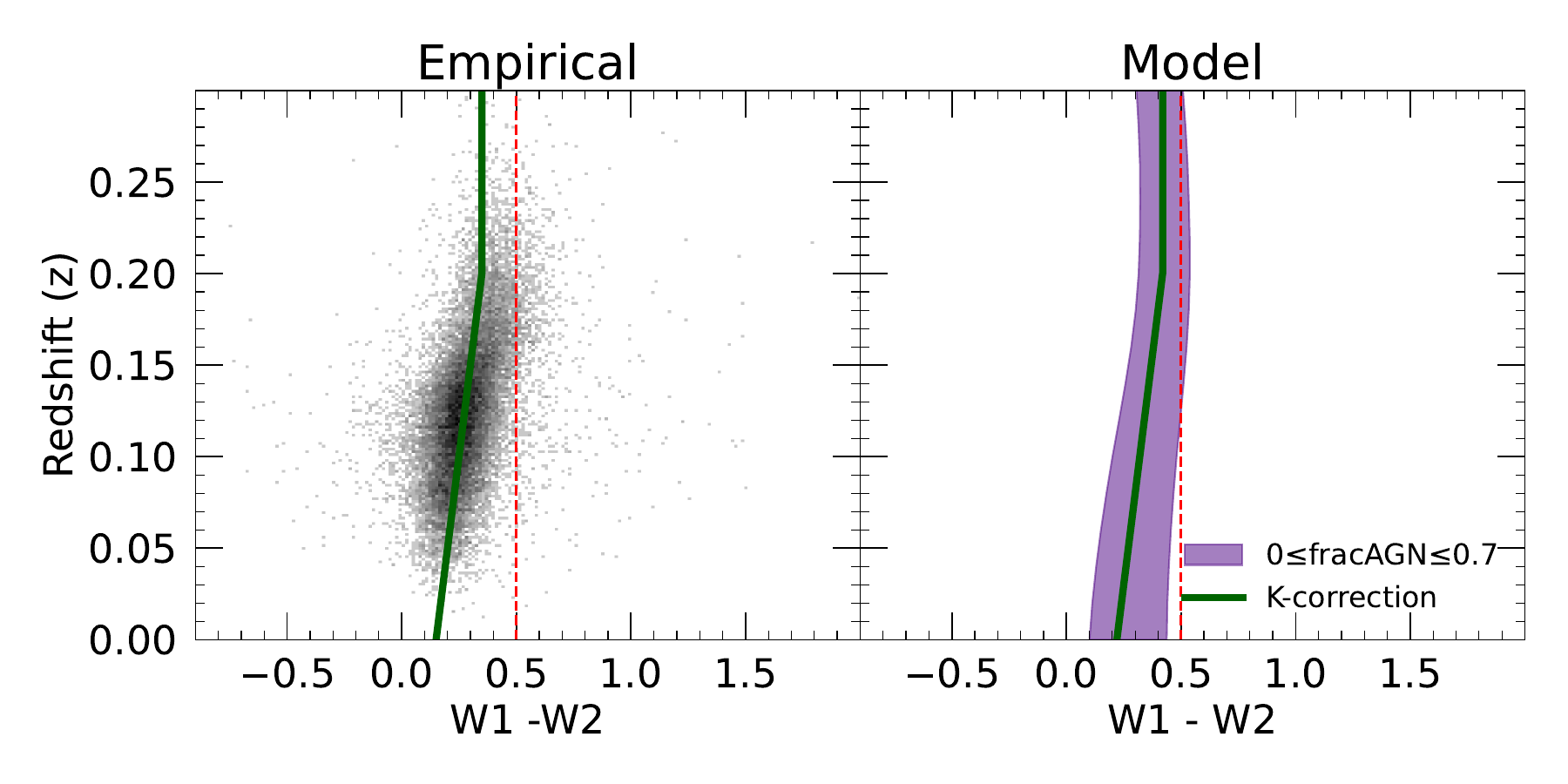} 
\caption{Derivation and validation of the K-correction. Left: The K-correction (green line) was obtained from the redshift vs.\ $W1-W2$ of SF galaxies. Right: The derived K-correction is consistent with the trends of the model colors for a range of AGN fractions and fixed metallicity, dust attenuation and SFH (tau decline rate).}
\label{fig:kcorr}
\end{figure}

\noindent This simple, piecewise linear K-correction was obtained empirically, by examining the redshift-color relation of SF galaxies. Specifically, we derive the correction only with galaxies having $\log$ SFR $>0.5$ (left panel of Figure \ref{fig:kcorr}), thus ensuring that galaxies with a small range of stellar metallicity, stellar mass and dust content are selected at all redshifts. This empirical K-correction is verified with the modeling (see Section \ref{sec:modeling} for details), choosing to fix the factors affecting the $W1-W2$ color (metallicity, dust attenuation, star formation history (SFH)). Modeling also indicates that the $W1-W2$ K-correction is similar for different levels of AGN contributions to the mid-IR, i.e., for different intrinsic mid-IR SED shapes, as demonstrated in the right panel of Figure \ref{fig:kcorr}. Most importantly, it is valid around $W1-W2=0.5$, where the selection cut is located.

After applying the K-correction (typically 0.1 mag, and no more than 0.2 mag), the right-hand panels of Figure \ref{fig:redshift} show that the color–redshift relation straightens significantly. Note that the color–redshift relation is not perfectly vertical, because the galaxies closer to upper redshift limit ($z=0.3$) where only the more massive galaxies are detectable, have higher metallicity, which affects the $W1-W2$ color to some extent. Importantly, the impact of K-correction is that the number of SF galaxies above the $W1-W2=0.5$ threshold is cut by a factor of three. Thus, applying the K-correction reduces the non-AGN contamination of the sample, although there are still SF galaxies with red and even very red colors, which we will investigate in Section \ref{subsec:res_sf_color}. In all the subsequent analysis, we adopt the K-corrected $W1-W2$ colors.  

\begin{figure}[ht!]
\centering
\begin{minipage}{0.45\textwidth}
    \centering
    \includegraphics[width=\linewidth, trim={0 0 0 0}, clip]{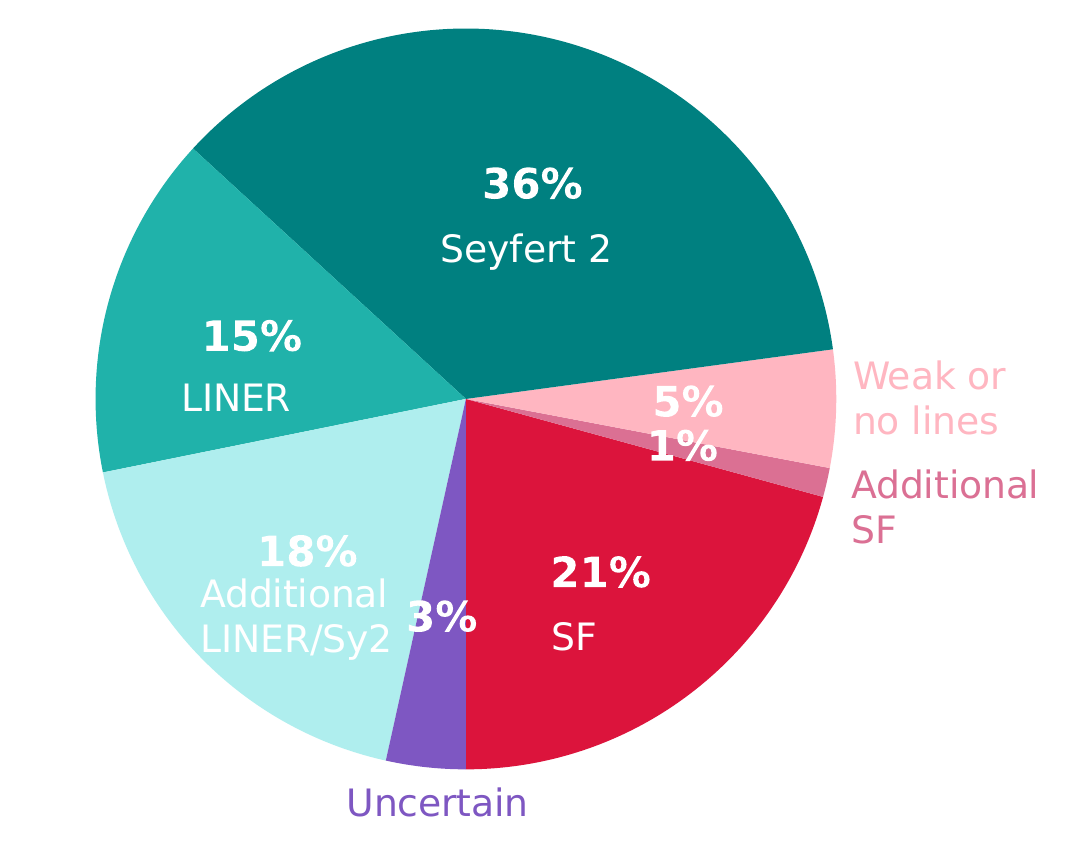}
\end{minipage}

\caption{Pie chart illustrating the fractional contributions of different galaxy classes (according to the emission lines) among the red galaxies ($(W1 - W2)_{[z=0]} > 0.5$). Around 30\% are not associated with emission-line AGNs.}
\label{fig:piechart}
\end{figure}

\subsection{Optimization of the WISE color cut} \label{subsec:res_optimization}

Our goal in this section is twofold: (1) test the effectiveness of different $W1-W2$ color cuts at identifying known AGNs and (2) evaluate how clean such a color cut is for preferentially selecting only AGNs.

As shown in the right panels of Figure \ref{fig:redshift}, only 13\% of all Sy2 galaxies in our sample exhibit $W1-W2$ colors greater than 0.5 (after the K-correction). The fraction of LINERs that gets selected as mid-IR AGN is even smaller, just $\sim$1\%. Percentages of red galaxies in other categories is given in Table \ref{table:sample B}.  

Furthermore, the piechart in Figure \ref{fig:piechart} presents the fractional contributions of each galaxy class among the red($W1-W2_{[z=0]} > 0.5$) galaxies. We see that the Seyfert 2s represent about a third of all red galaxies. If we combine the contributions of other likely AGN categories (LINERs and additional Sy2 and LINERs), we reach $\sim$70\%. The remaining $\sim$ 30\% are not AGNs according to the emission lines, so they either represent potential contamination among the real AGNs, or actual AGNs not identified by the BPT diagram, as discussed in Section \ref{sec:intro}. 


\begin{figure}[ht!]
\centering
\includegraphics[width=0.5\textwidth]{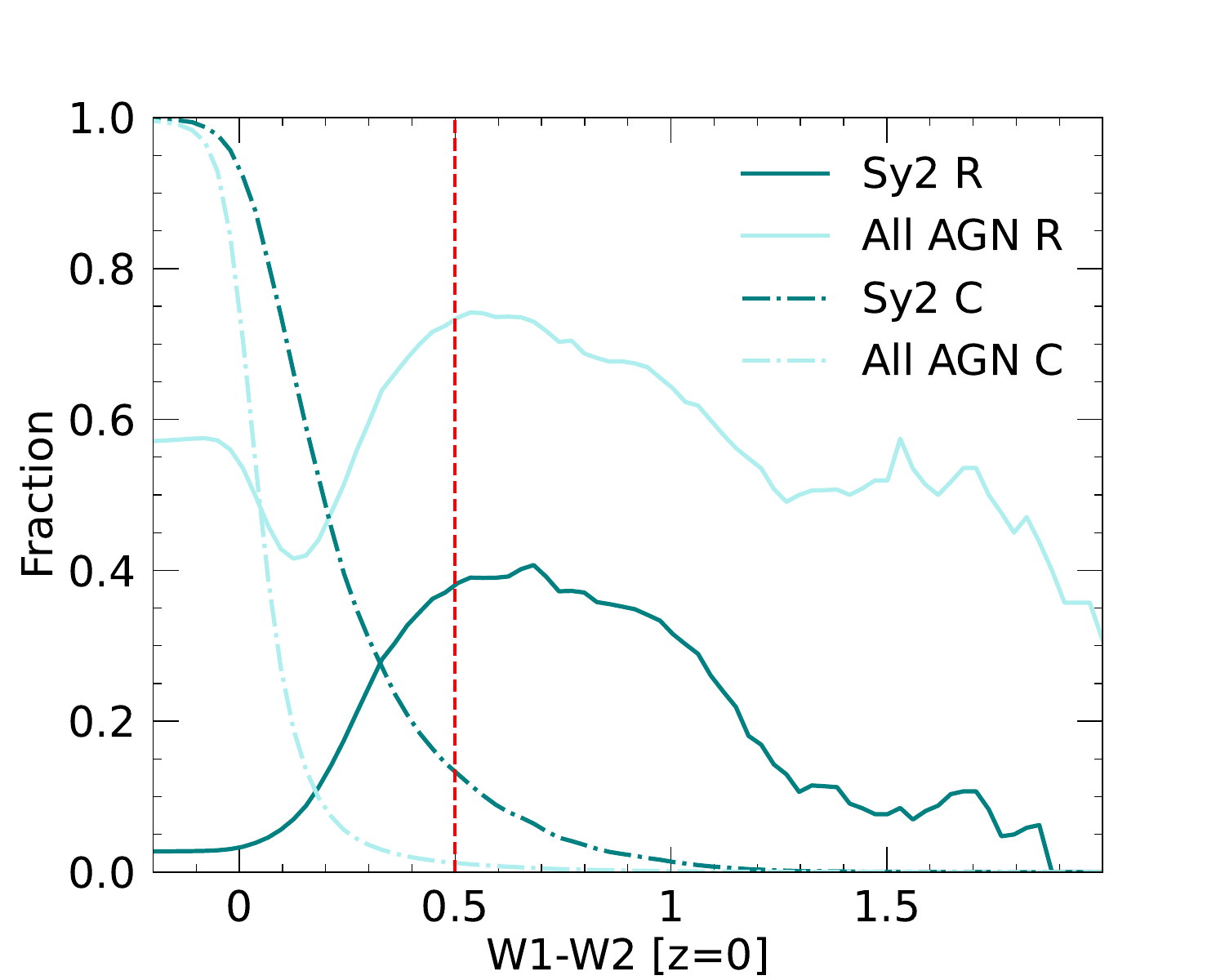} 
\caption{Reliability (R, solid line) and completeness (C, dashed line) as a function of the $W1-W2$ threshold for 1) Sy2 galaxies (dark green) and 2) combined Sy2, LINER and ``Additional LINER or Sy2" galaxies (light green). The reliability is defined as the fraction of galaxies above the threshold that are emission-line AGNs (Sy2 or Sy2, LINER and ``Additional LINER or Sy" combined), whereas the completeness is defined as the fraction of the emission-line AGN population (Sy2 or  Sy2, LINER and ``Additional LINER or Sy'' combined) recovered above the threshold. A vertical dashed line at $W1-W2=0.5$ is included to highlight our nominal color cut. This cut is near maximum reliability.}
\label{fig:rc}
\end{figure}

To assess whether changing the $W1-W2$ threshold from 0.5 to another value would yield a higher completeness of selecting Sy2 (currently only 13\%), while simultaneously not significantly increasing the contribution of non-AGN categories among red galaxies, we performed a reliability and completeness analysis. We define completeness as the fraction of BPT-AGNs that we identify as mid-IR AGN using a color cut. Reliability is defined as the fraction of mid-IR color-selected objects (red galaxies) that are BPT-AGNs, i.e., it quantifies how ``contaminated'' the red sample is by non-AGN galaxies (at least according to the emission lines).

We applied the completeness/reliability analysis to two BPT-AGN populations: the Sy2 galaxies alone (i.e., indisputable AGN), and a wider BPT-AGN population that includes Sy2, LINERs and ``Additional LINER or Sy2'' category. For each population we calculated the completeness and reliability as a function of $W1-W2$ threshold. Figure \ref{fig:rc} shows the resulting reliability (solid line) and completeness (dash dot line) curves for both populations. As we have seen, the nominal $W1-W2$ threshold of 0.5 recovers approximately 13\% of all Sy2 and less than 1\% of a broader population of AGN. At the same threshold, about 40\% of the galaxies selected by the color cut are Sy2 and about 70\% are AGNs of all types. Increasing the threshold does not improve the reliability (it is already near the peak at 0.5), while it lowers the completeness, because fewer Sy2s or AGNs in general would meet the stricter color cut. Conversely, lowering the threshold increases completeness (e.g., it would be around 40\% for Sy2 at $(W1 - W2)_{[z=0]} = 0.25$), but reduces reliability due to the increased contamination (e.g., at $(W1 - W2)_{[z=0]} > 0.25$ less than 20\% are Sy2).

To conclude, although the $(W1-W2)_{[z=0]}=0.5$ color cut does not yield high completeness for Sy2s or the combined AGN population, it does offer a good compromise by being near the peak of the reliability. Lowering it improves completeness but increases contamination. We retain this cut for the rest of the analyses.

The analysis presented in this section highlights two key points:
\begin{enumerate}
    \item The WISE selection criterion only recovers only 13\% of the Sy2 galaxies and 1\% of LINER galaxies. This raises the question of why a significant fraction of AGN (especially Sy2, as bona-fide AGNs) are not recovered by the color cut.
    \item The application of K-correction eliminated a significant number of contaminants. However, despite that, 0.6\% of SF galaxies pass the color cut, which make up about 20\% of the red sample (presumably mid-IR AGN). This prompts the question of whether these systems, despite being classified as SF by the BPT diagram, host an AGN.
\end{enumerate}

The following sections address these two issues in detail. 

\subsection{Why Do Most Emission-line AGNs Fail the WISE Color Cut?}
\label{sec:modeling}

To investigate why the majority of the AGN population in our sample do not get selected by the $W1-W2>0.5$ color cut, we will model this color using the combination of stellar and dust emission models using the Code Investigating GALaxy Emission (CIGALE) \citep{cigale}. In particular, we want to assess how the variations in the AGN fraction affects the $W1-W2$ color cut as an AGN diagnostic tool. 

\begin{figure}[h!]
\centering
\includegraphics[width=0.5\textwidth, trim={0 0 0 30}]{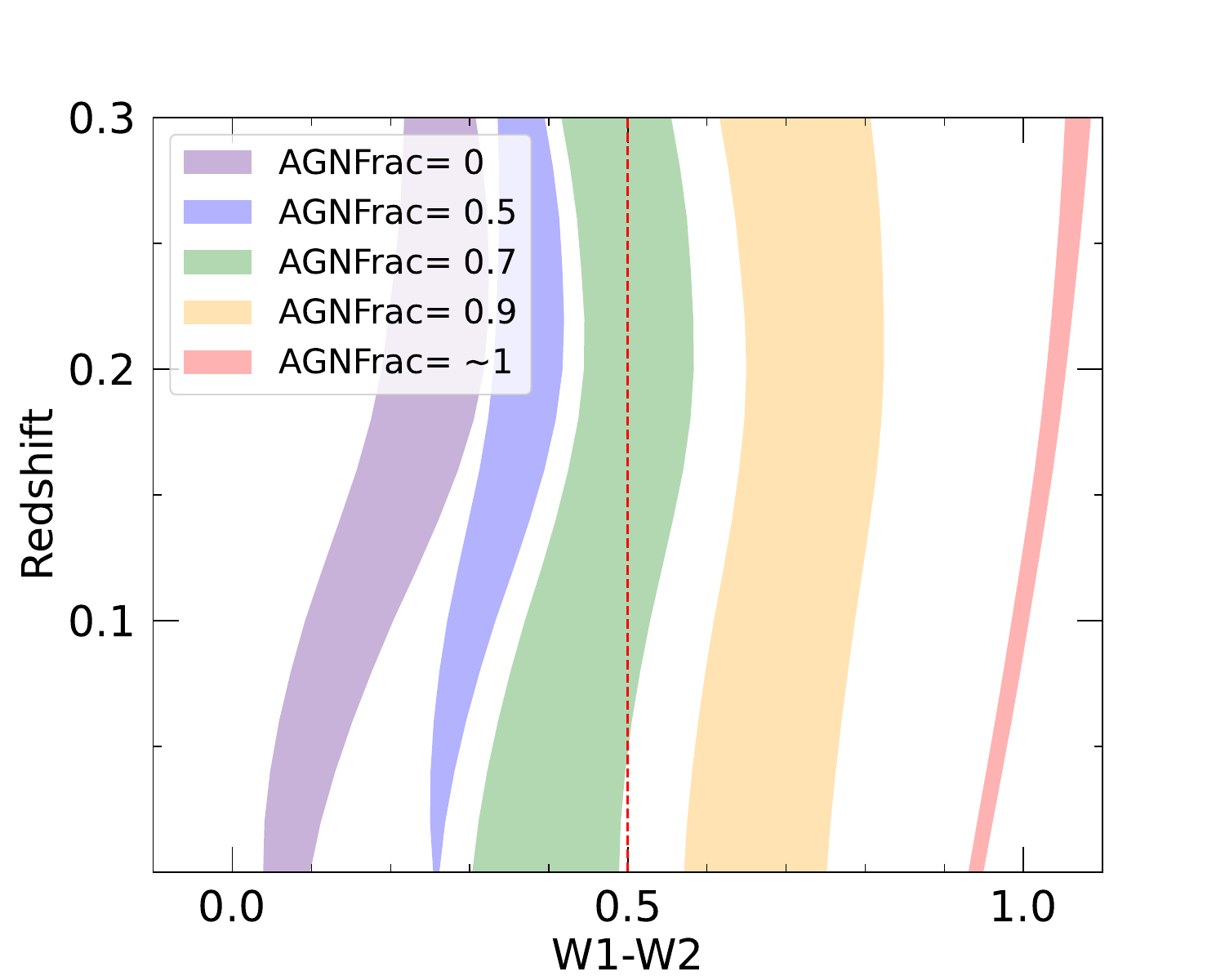} 
\caption{Model observed-frame $W1-W2$ color for varying AGN fractions. Each colored band corresponds to a fixed AGN fraction (AGN contribution to total IR luminosity) indicated in legend, with the band spread arising from a spread in other parameters (tau decline rates, dust attenuation, and metallicity). The figure illustrates that only models with high AGN contributions (\texttt{fracAGN} $\gtrsim 0.75$) attain $(W1-W2)_{[z=0]} >$ 0.5.}
\label{fig:models}
\end{figure}

The following modules are used to model the galaxy and AGN components:
\begin{enumerate}
    \item Star formation history: \texttt{sfh2exp} --- we use a simple exponentially declining SFH with the onset at 10 Gyr in the past and decline times between 1.3 and 2.4 Gyr.
    \item The stellar population model: \texttt{bc03} --- based on the simple stellar population library of \cite{2003MNRAS.344.1000B} with subsolar, solar and supersolar metallicities.
    \item Dust attenuation: \texttt{dust\_calzleit} --- based on the \cite{Noll2009A&A...507.1793N} and  \cite{2018ApJ...859...11S} generalization of the \cite{2000ApJ...533..682C} attenuation curve, expanded to far UV using \cite{2002ApJS..140..303L} and effective $E(B-V)$ from 0.07 to 0.22.
    \item Dust emission: \texttt{dale2014} --- dust emission templates from \citet{2014ApJ...784...83D}
\end{enumerate}

The \citet{2014ApJ...784...83D} templates are beneficial due to their simplicity for modeling both the stellar and AGN heated dust. They include the parameter \texttt{alpha} to characterize the shape of the IR SED produced by the dust heated by the stellar component. The AGN is parameterized by the \texttt{fracAGN} --- the fractional contribution of the AGN to the total (stellar + AGN) infrared luminosity. Parameter \texttt{alpha}  has no effect on $W1$ and $W2$, so we keep it fixed. We vary \texttt{fracAGN}  between 0 and nearly 1.

Model sets were computed for different redshifts with variations in the tau decline rate of the exponential SFH, dust attenuation normalization ($E(B-V)$) and stellar metallicity. Other parameters were kept fixed. The resulting models, shown in Figure \ref{fig:models}, illustrate the trend in the observed $W1-W2$ color with redshift for different AGN fractions. First, we see that the observed colors depend on redshift in a way that agrees with the K-correction derived in Section \ref{subsec:res_k_correct}. Next, we find that increasing either the dust reddening or the stellar metallicity results in redder $W1-W2$ color. However, these changes are small compared to the effect of changing the AGN fraction and were incorporated into the shaded contour. The AGN fraction has the strongest influence: only models with \texttt{fracAGN} $\gtrsim 0.75$ produce $(W1-W2)_{[z=0]} >$ 0.5. Lower AGN fractions, even with favorable dust or stellar parameters, do not reach this color cut. Since only the galaxies with very high AGN fraction cross the color cut threshold, it is not surprising that the AGN yield from the color cut is not very high. Presumably, only the more luminous and/or more obscured AGN get selected by the color cut. We will explore the physical factors that lead to high AGN fraction in the following section.

\begin{figure*}[htbp]
    \centering
    \begin{minipage}{0.32\textwidth}
        \centering
        \includegraphics[width=\linewidth]{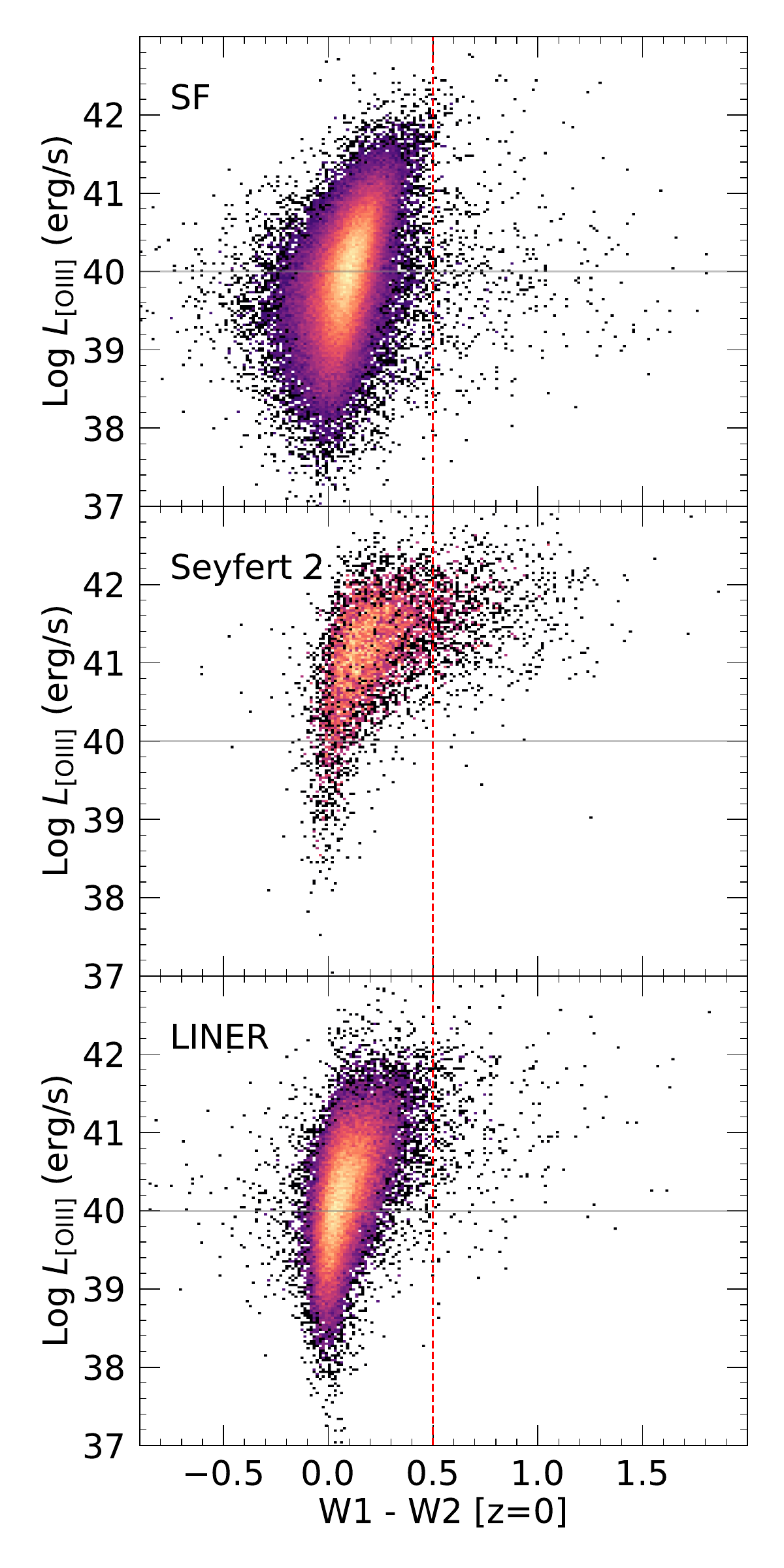}
    \end{minipage}\hfill
    \begin{minipage}{0.32\textwidth}
        \centering
        \includegraphics[width=\linewidth]{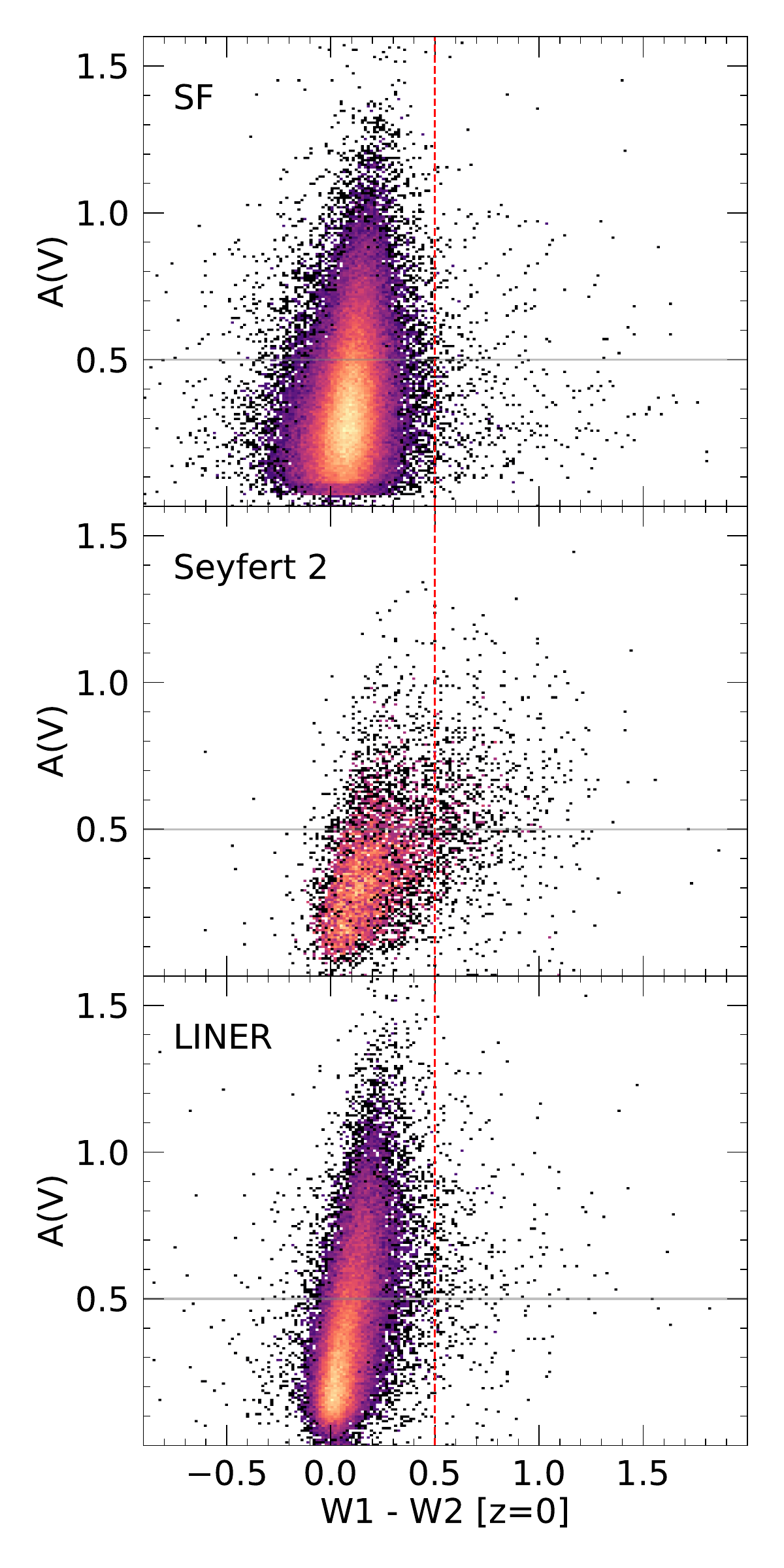}
      \end{minipage}\hfill
        \begin{minipage}{0.32\textwidth}
        \centering
        \includegraphics[width=\linewidth]{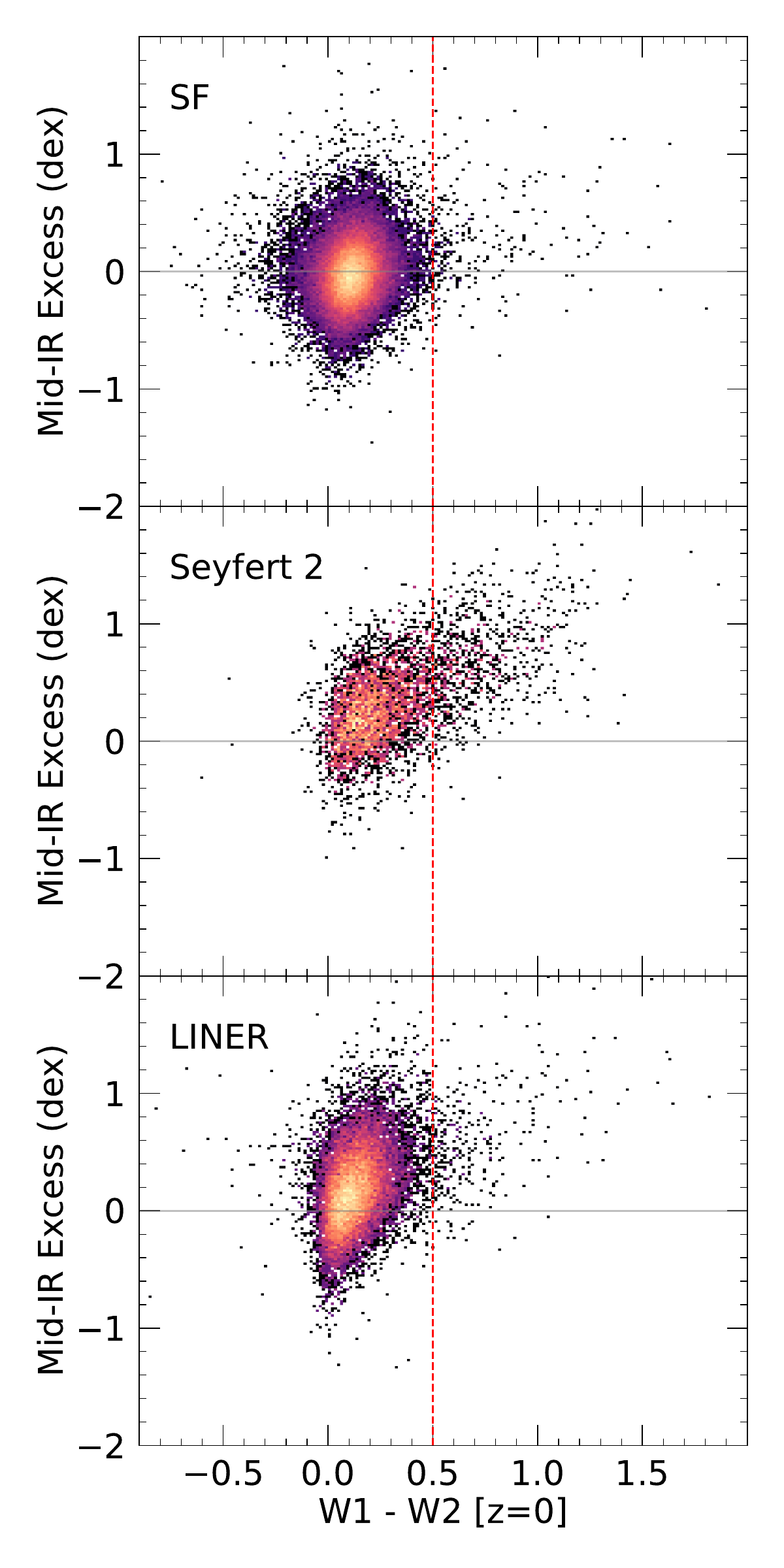}
    \end{minipage}\hfill
    \caption{Galaxy parameters as a function of $(W1-W2)_{[z=0]}$ for SF, Sy2 and LINER galaxies. \textit{Left}: Dust-corrected $\mathrm{[OIII]}$ luminosity. Luminous AGNs (Sy2s/LINERs) are not exclusive to red colors, indicating that high AGN luminosity is a necessary, though not sufficient, condition for red mid-IR colors. Red LINER galaxies occupy a similar $L_{\rm [OIII]}$ regime as Sy2s. \textit{Middle}: Global dust attenuation $(A(V))$ vs.\ $(W1-W2)_{[z=0]}$. Red Sy2 galaxies tend to show higher attenuation than blue Sy2s, consistent with dust reprocessing of AGN emission. \textit{Right}: Mid-Infrared Excess, defined as the ratio of $W4$-derived star formation rate to UV/optical star formation rate, vs.\ $(W1-W2)_{[z=0]}$. Both red Sy2 and LINER galaxies show significant positive mid-IR excess, supporting AGN contribution to warm dust emission. Red SF galaxies show a modest but predominantly positive excess above $(W1-W2)_{[z=0]} > 0.5$ suggesting non-stellar heating as well. Horizontal lines are included to help compare the distributions between the categories. }
    \label{fig:drivers1}
\end{figure*}
  
\subsection{Identifying the Physical Origin of Red WISE Colors}
\label{subsec:redwise}

To identify the processes responsible for producing red $W1-W2$ colors in our galaxies, we study several observed properties, including ionized gas emission, star formation activity, line kinematics, and dust attenuation. With this approach, we hope to distinguish the contributions from AGN activity and from stellar processes. In particular, the aim of this section is to:

\begin{enumerate}
\item Investigate which factors drive red $W1-W2$ colors (and hence high AGN fractions) in Sy2 galaxies.
\item Evaluate whether the  $W1-W2$ red LINERs represent genuine AGN, given the ongoing debate surrounding the AGN nature of LINERs.
\item Assess whether there is any underlying AGN contribution in $W1-W2$ red SF galaxies, which according to the BPT diagram, and after applying the K-correction, are not expected to host an AGN.
\end{enumerate}

\subsubsection{Which factors control the AGN fraction parameter in Seyfert 2s?
}
Modeling in Section \ref{sec:modeling} indicated that achieving $(W1-W2)_{[z=0]} > 0.5$ required AGN fractional contribution to IR luminosity (\texttt{fracAGN}) of at least $\sim$0.75. However, the AGN fraction is just a phenomenological parameter and it does not tell us why it is low or high.

The first potential driver is the AGN luminosity itself (e.g., \citealt{Blecha2018MNRAS.478.3056B}). Unfortunately, the bolometric luminosity of an AGN cannot be easily determined directly. An often used proxy is the (dust corrected) luminosity of [OIII]$\lambda$5007 line \citep{Kauffmann2003}. As shown in the middle left panel of Figure \ref{fig:drivers1}, we find that Sy2 galaxies with red $W1-W2$ colors consistently show [OIII] luminosities ($L_{[\mathrm{OIII}]}$) higher than $10^{40} \mathrm{\ erg\,s^{-1}}$, with a notable absence of red colors in the lower-luminosity regime ($L_{[\mathrm{OIII}]} < 10^{40} \mathrm{erg\,s^{-1}}$). However, high $L_{[\mathrm{OIII}]}$ does not always lead to red colors, so it seems to be a necessary, but not a sufficient condition. Another possibility is that the red $W1-W2$ color is driven by AGN luminosity, but normalized by host property. We have thus explored the trends of $L_{[\mathrm{OIII}]}$ normalized by $M_*$ and by $M_{\mathrm{bulge}}$, but these have not revealed them as being sufficient for red color. 


High AGN luminosity does not automatically mean that much of that luminosity is absorbed by the dust. To try to assess the latter, we also look at the trends of the K-corrected $W1-W2$ color and dust attenuation $A_V$. As seen from the middle panel of Figure \ref{fig:drivers1}, red Sy2 galaxies show increasing $W1-W2$ colors with increasing attenuation. This is consistent with obscured AGN where the nuclear dust reprocesses emission from the accretion disk. However, as in the case of [OIII] luminosity, the correlation is not such that higher $A_V$ values definitely lead to red $W1-W2$ color. Because these attenuation values are global and pertain to the stellar continuum, they may only loosely reflect the true nuclear obscuration. We thus performed a similar analysis using the nebular attenuation derived from the Balmer decrement, which tells us about the attenuation of the NLR, but the results are qualitatively similar. Likewise, none of the other physical properties we explored is solely associated with red colors.

To conclude, the physical drivers behind high AGN fraction parameter, and therefore the red $W1-W2$ color, are consistent with some combination of intrinsically high AGN luminosity and the amount of dust, or the dust opacity, in the nuclear region. However, assessing this relation via observationally accessible parameters is challenging.

\begin{figure*}[htbp]
  \centering
    \begin{minipage}{0.32\textwidth}
        \centering
        \includegraphics[width=\linewidth]{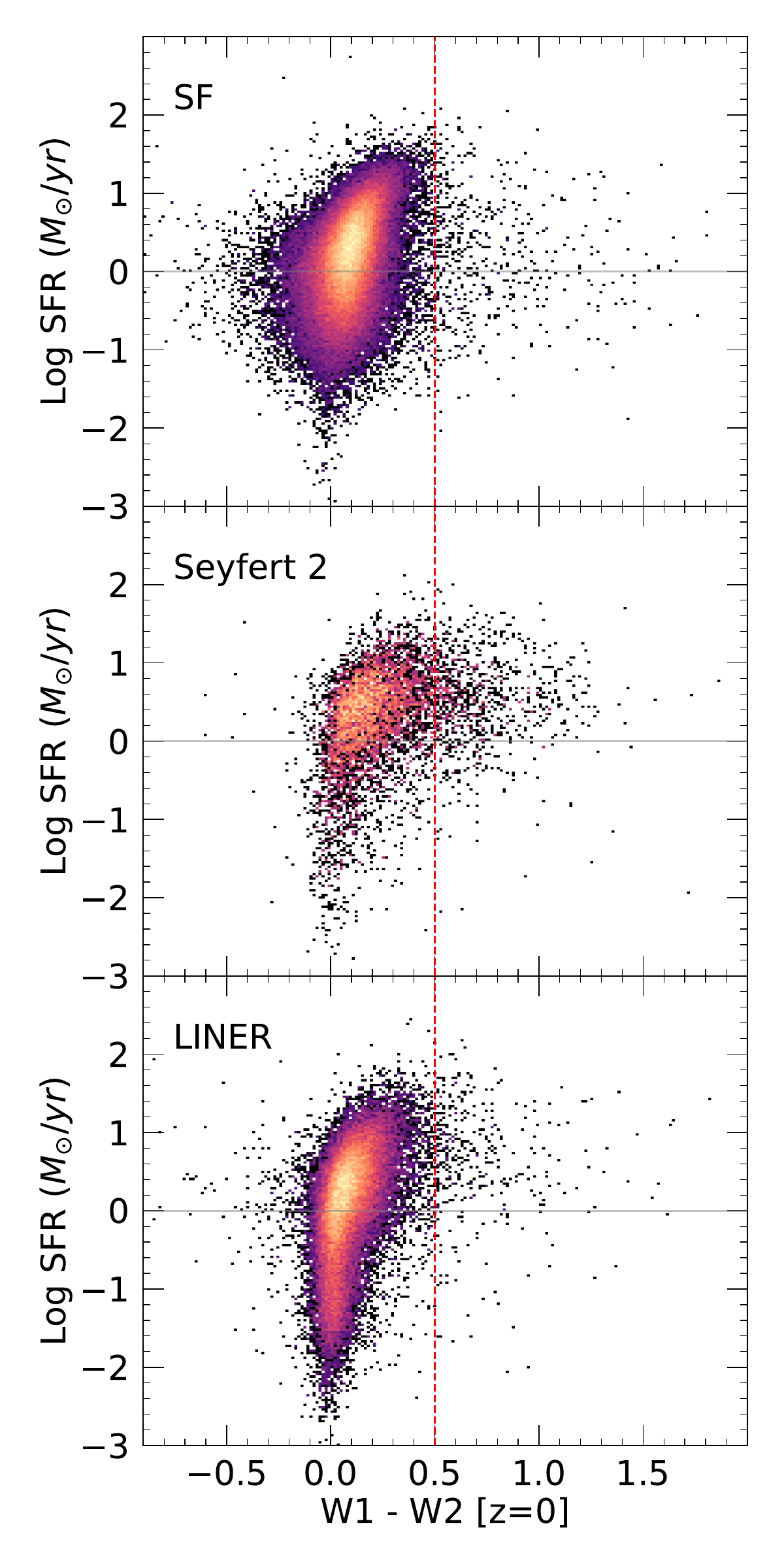}
    \end{minipage}
    \begin{minipage}{0.32\textwidth}
        \centering
        \includegraphics[width=\linewidth]{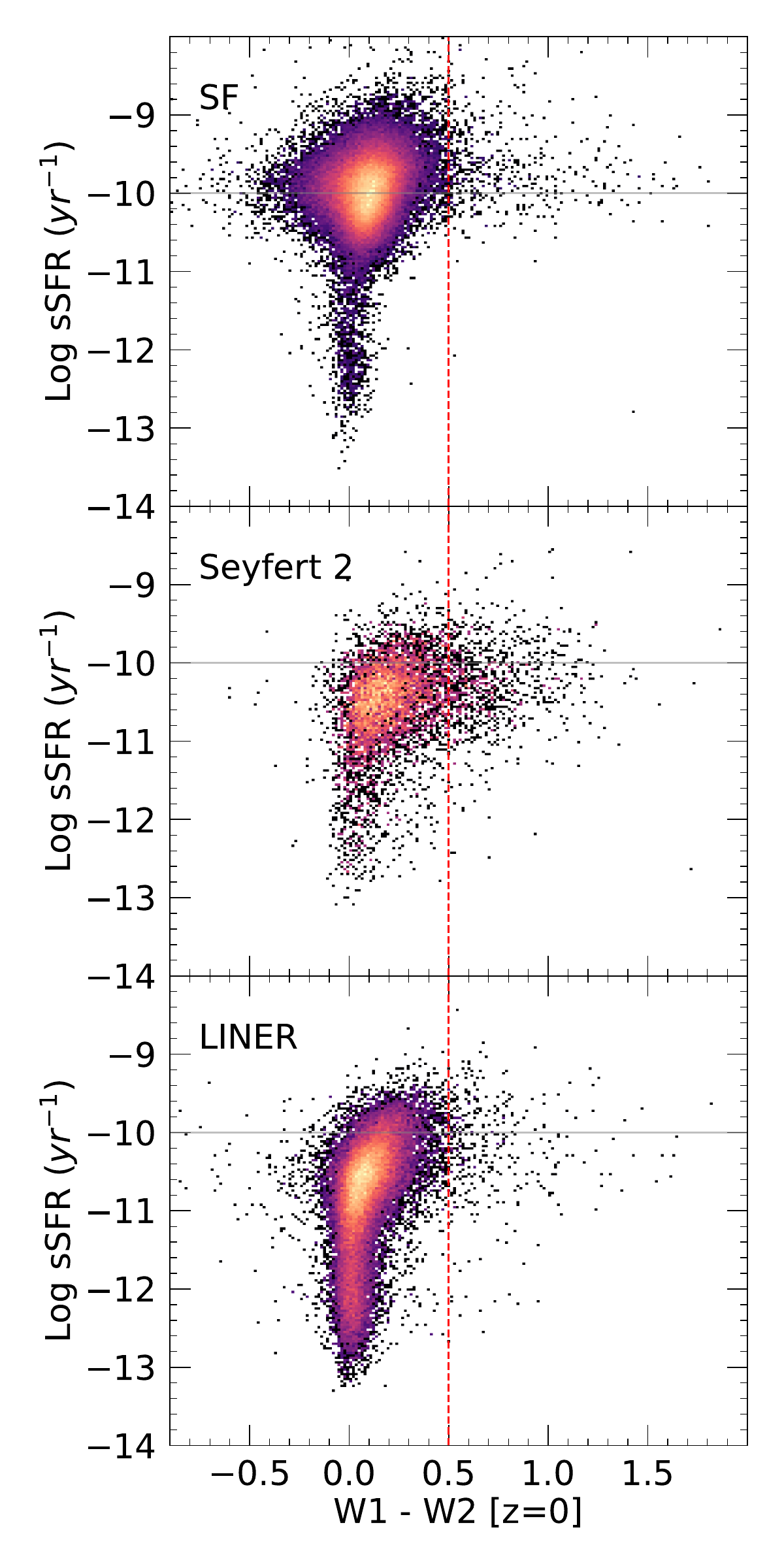}
      \end{minipage}\hfill
    \caption{Star Formation Rate (SFR) and specific Star Formation Rate (sSFR) as a function of $(W1-W2)_{[z=0]}$ color for SF, Sy2 and LINER galaxies. \textit{Left}: SFR vs.\ $(W1-W2)_{[z=0]}$. Red SF galaxies span a wide range of SFR. \textit{Right}: sSFR, defined as SFR normalized by stellar mass, vs.\ $(W1-W2)_{[z=0]}$. Red SF galaxies occupy a broad range of sSFR values and are not preferentially found at the highest sSFRs.}
    \label{fig:drivers2}
\end{figure*}

\subsubsection{Are red LINERs true AGNs?}

Whether the emission lines in LINERs arise from AGN activity or stellar sources, such as old stellar populations, has been debated for years, with some results suggesting that all LINERs are AGN and others that none are (\citealt{heckman1980A&A....87..152H, Ho1993ApJ...417...63H, 1994A&A...292...13B, 2008MNRAS.391L..29S, 2010MNRAS.402.2187S, 2011MNRAS.413.1687C, 2012ApJ...747...61Y, 2013A&A...558A..43S, 2016MNRAS.461.3111B, 2016A&A...588A..68G, 2017A&A...599A.141J, 2017MNRAS.466.3217Z, 2019AJ....158....2B}). Here we can assess this question in particular for the case of red LINERs, which, according to their colors, should contain a mid-IR AGN. 

LINERs on average occupy a lower [OIII] luminosity space compared to Sy2 galaxies, often below $L_{[\mathrm{OIII}]} \approx 10^{41} \mathrm{\ erg \ s^{-1}}$ (bottom and middle left panels of Figure \ref{fig:drivers1}). However, the red LINERs are found among the galaxies with similar [OIII] luminosities as Sy2s, mostly above $10^{40} \mathrm{\ erg \ s^{-1}}$. We note that the similarity of [OIII] luminosities between red LINERs and red Sy2s is more pronounced with dust-corrected [OIII] luminosities. Without the dust correction, the [OIII] luminosity of LINERs tend to be about 1 dex lower than that of Sy2s. In other words, LINERs are subject to higher nebular dust attenuation than Sy2s, but other than that the reason for their red colors seems to be the same as for Sy2s. 

Next we turn our attention to the mid-IR excess. We define the mid-IR excess as the ratio between the IR-derived SFR and the dust-corrected UV/optical SFR:

\begin{equation} \label{eq:midirexcess}
\text{Mid-IR excess} = \frac{\mathrm{SFR_{IR}}}{\mathrm{SFR_{UV/Opt\, SED}}}
\end{equation}

\noindent The IR SFR is calculated from IR luminosity using a simple conversion factor and thus assuming that the entire IR luminosity (inferred from $W4$, or $W3$ if $W4$ is not detected, using the non-AGN dust templates; see \citealt{gswlc-1} for details) originates from the dust heated by stars. Since $W4$ could in reality include AGN contribution, the SFR we derive in this way will be artificially boosted in such cases. On the other hand, the SED SFR in the denominator comes from the UV/optical SED fitting (no IR), and provides the true dust-corrected estimate (from the SED shape) of the SFR (unbiased by any AGN).  In systems where dust heating is dominated by star formation, the IR SFR is expected to equal that predicted from dust-corrected UV/optical SFR (excess $=1$), whereas a significant positive mid-IR excess would indicate additional dust heating beyond that expected from star formation alone, which may be coming from AGN heated dust. 


In the middle right panel of Figure \ref{fig:drivers1} we see a strong excess for red Sy2 galaxies. The excess is typically below 1 dex, but occasionally extends to 2 dex. Of all the quantities explored in this study the mid-IR excess most strongly correlates with the red $W1-W2$ color. Interestingly, the mid-IR excess is on average positive even for Sy2 galaxies that are not red, suggesting it is a more sensitive tracer of AGN dust heating than $W1-W2$. Looking now at the LINERs, we see that red LINERs exhibit a strong mid-IR excess, similar to Sy2 galaxies, as shown in the bottom right panel of Figure \ref{fig:drivers1}. And like Sy2 galaxies, LINER galaxies on average show a mild mid-IR excess even at $(W1-W2)_{[z=0]} < 0.5$. It should be pointed out that this mid-IR excess is a relatively noisy measure, so while it is useful to characterize the presence or absence of AGNs in bulk, it is not possible to use it to robustly select AGNs in individual galaxies. 

\begin{figure*}[htbp]
  \centering
    \begin{minipage}{0.32\textwidth}
        \centering
        \includegraphics[width=\linewidth]{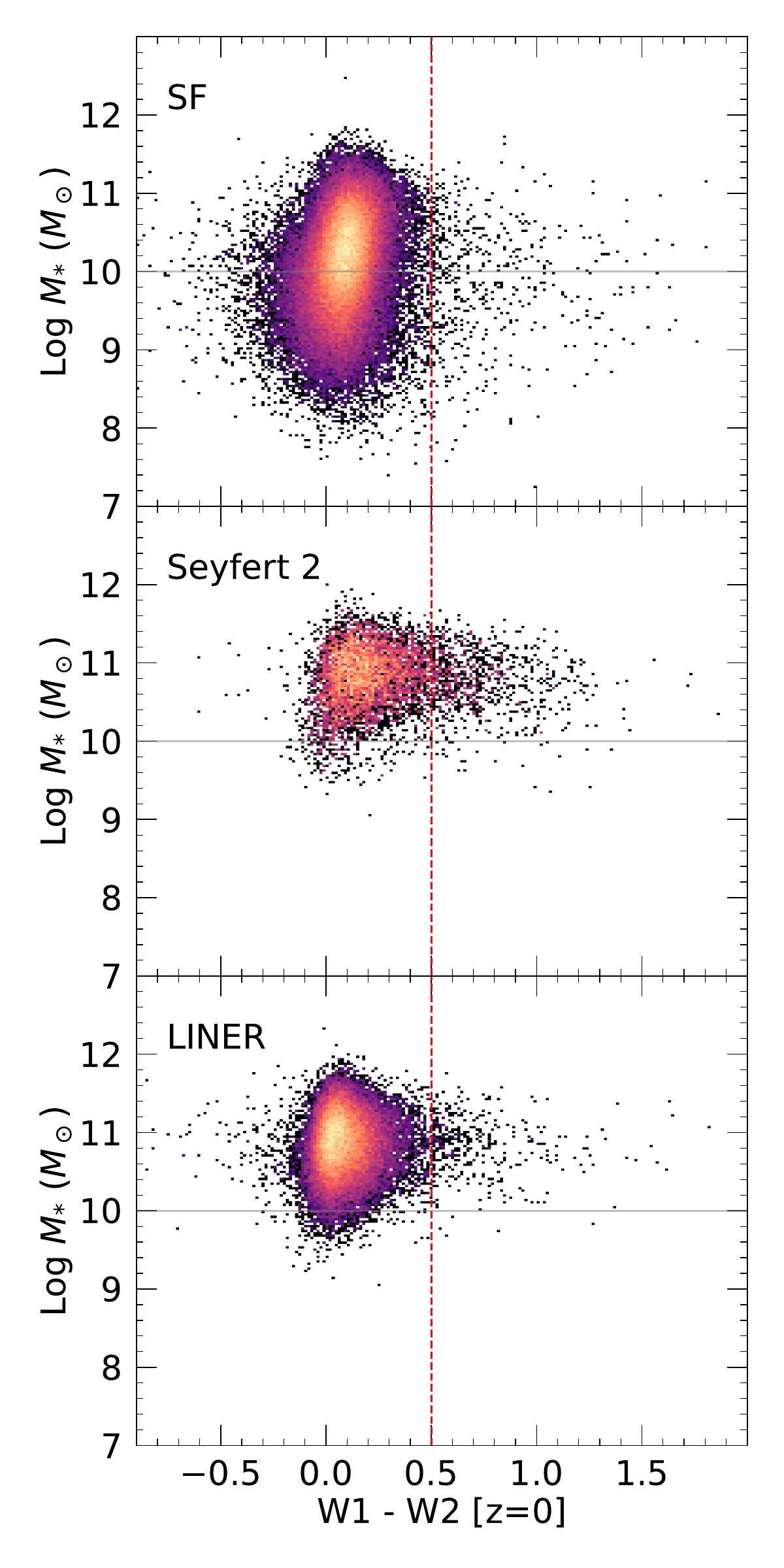}
    \end{minipage}
    \begin{minipage}{0.32\textwidth}
        \centering
        \includegraphics[width=\linewidth]{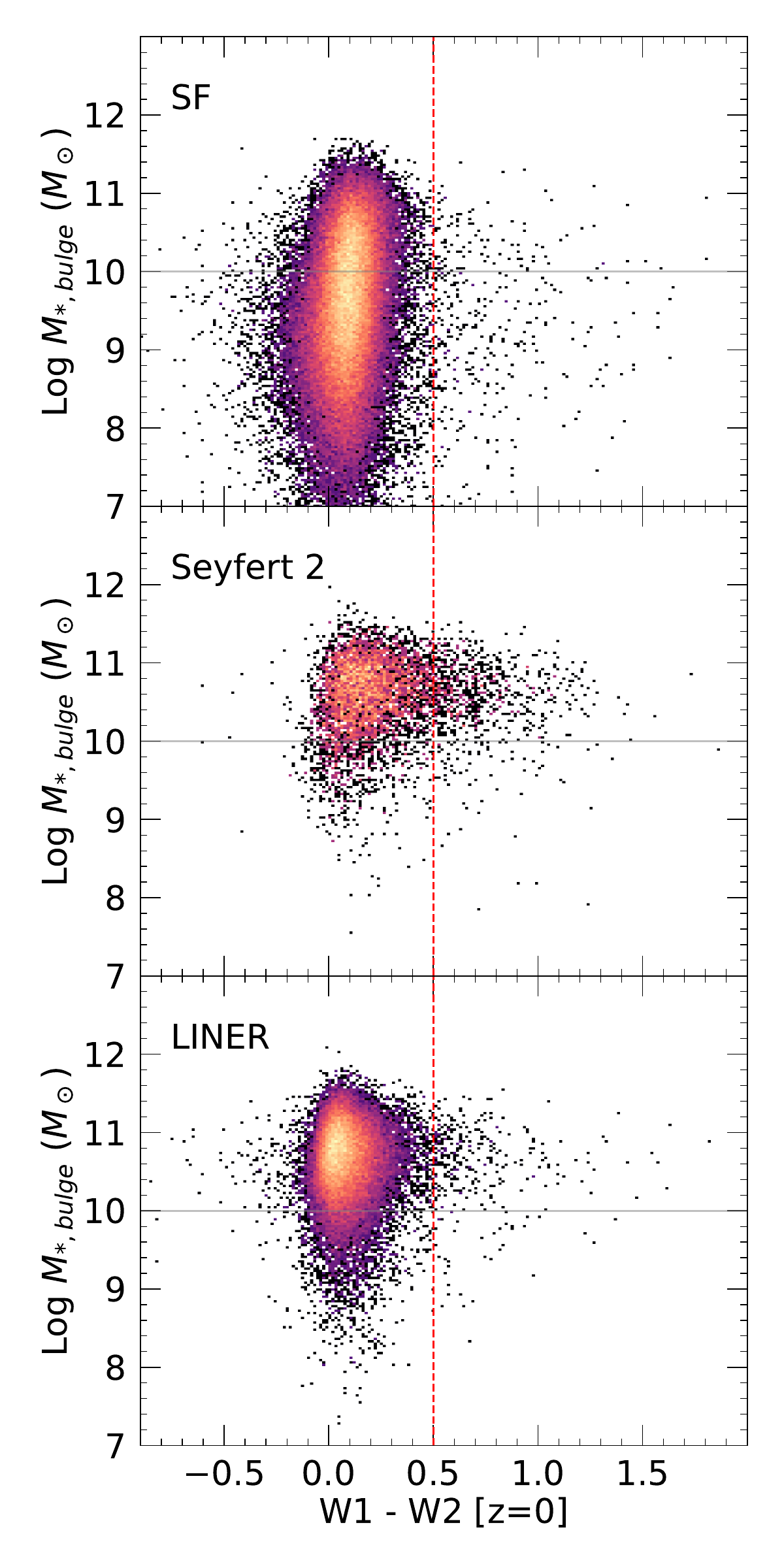}
      \end{minipage}
            \begin{minipage}{0.32\textwidth}
        \centering
        \includegraphics[width=\linewidth]{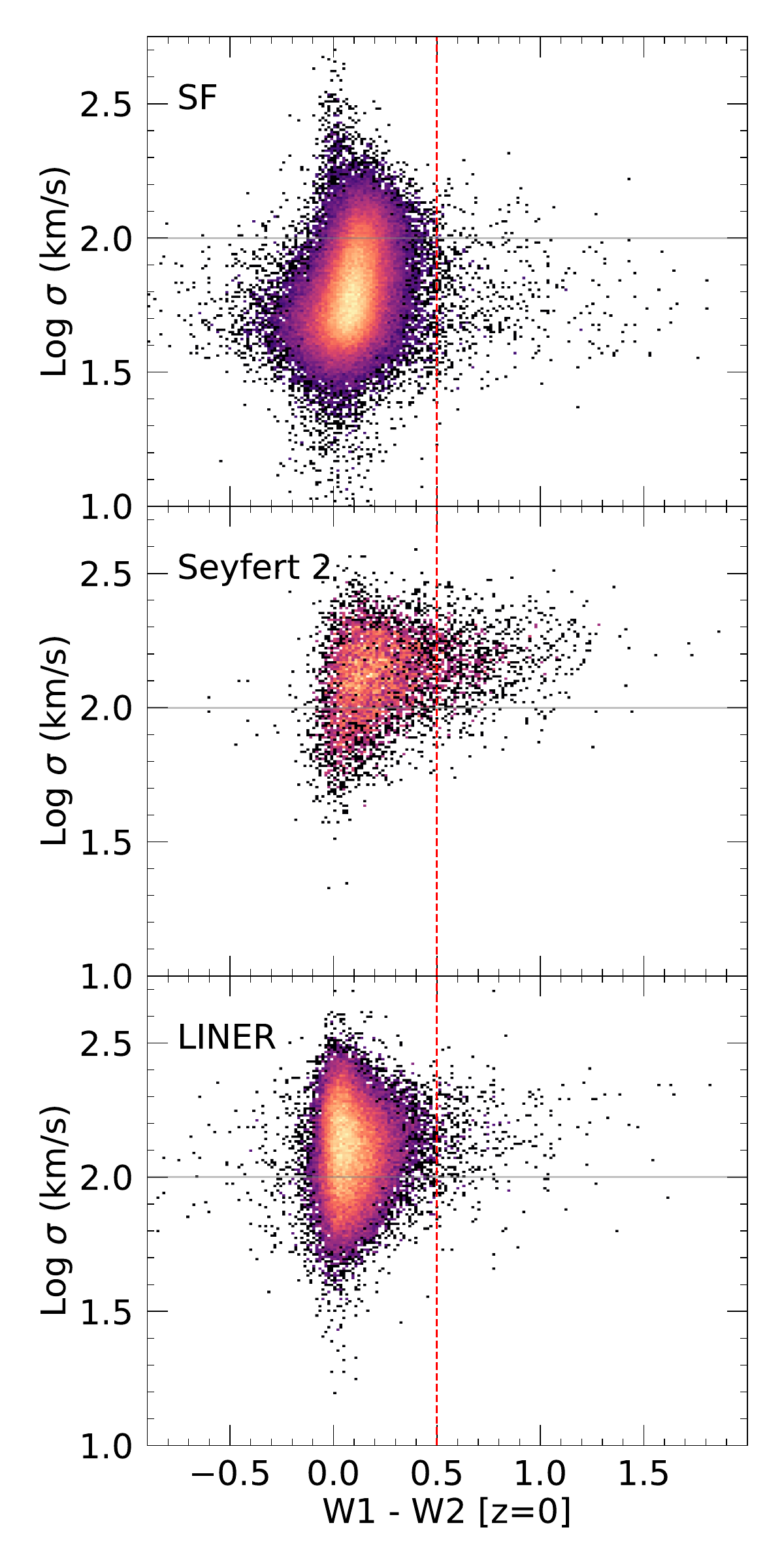}
    \end{minipage}\hfill
    \caption{Mass-related properties as a function of $(W1-W2)_{[z=0]}$ color for SF, Sy2 and LINER galaxies. \textit{Left}: Stellar mass ($\mathrm{M_*}$) vs.\ $(W1-W2)_{[z=0]}$. Red Sy2 and red LINER galaxies are exclusively found in galaxies with high stellar mass, whereas red SF galaxies span a broader mass range. \textit{Middle}: Bulge stellar mass ($\log \ M_{\mathrm{bulge}}$) vs.\ $(W1-W2)_{[z=0]}$. Red Sy2 and LINER galaxies are mostly in massive bulges ($> 10^{10} \ \mathrm{M_\odot}$), while most red SF galaxies have lower bulge mass in comparison. \textit{Right}: Forbidden-line velocity dispersion ($\sigma$) vs.\ $(W1-W2)_{[z=0]}$. Red Sy2 and LINER galaxies exhibit high velocity dispersions ($\sigma> 100 \mathrm{\ km\, s^{-1}}$), whereas red SF galaxies largely fall below this threshold.  Taken together, these trends show a host galaxy dichotomy in which mid-IR selected BPT-AGNs reside in massive bulges (bulges with high velocity dispersions), while red BPT-SF galaxies preferentially inhabit lower mass bulges.} 
    \label{fig:drivers3}
\end{figure*}

Finally, we note LINER and Sy2 galaxies have similarly massive hosts ($> 10^{10} \ M_\odot$), whether considering the entire galaxies or just their bulges (their $B/T$ is typically near unity), and, consequently, similarly high velocity dispersions, extending past 100 $\mathrm{\ km\, s^{-1}}$ (middle and bottom panels of Figure \ref{fig:drivers3}.) This suggests that both LINER and Sy2 galaxies have comparable gravitational potentials and likely host black holes of similar masses.

From these analyses we conclude that red LINERs are almost certainly AGN, and, as in the case of Sy2s, they are associated with more luminous AGN and are situated in more massive bulges. Furthermore, the majority of blue LINERs might be AGN as well, judging by their mid-IR excess, albeit it is (as expected) more moderate than for red LINERs.

\begin{figure}[htb!]
\centering
\hspace*{-0.8cm}\includegraphics[width=1.1\columnwidth]{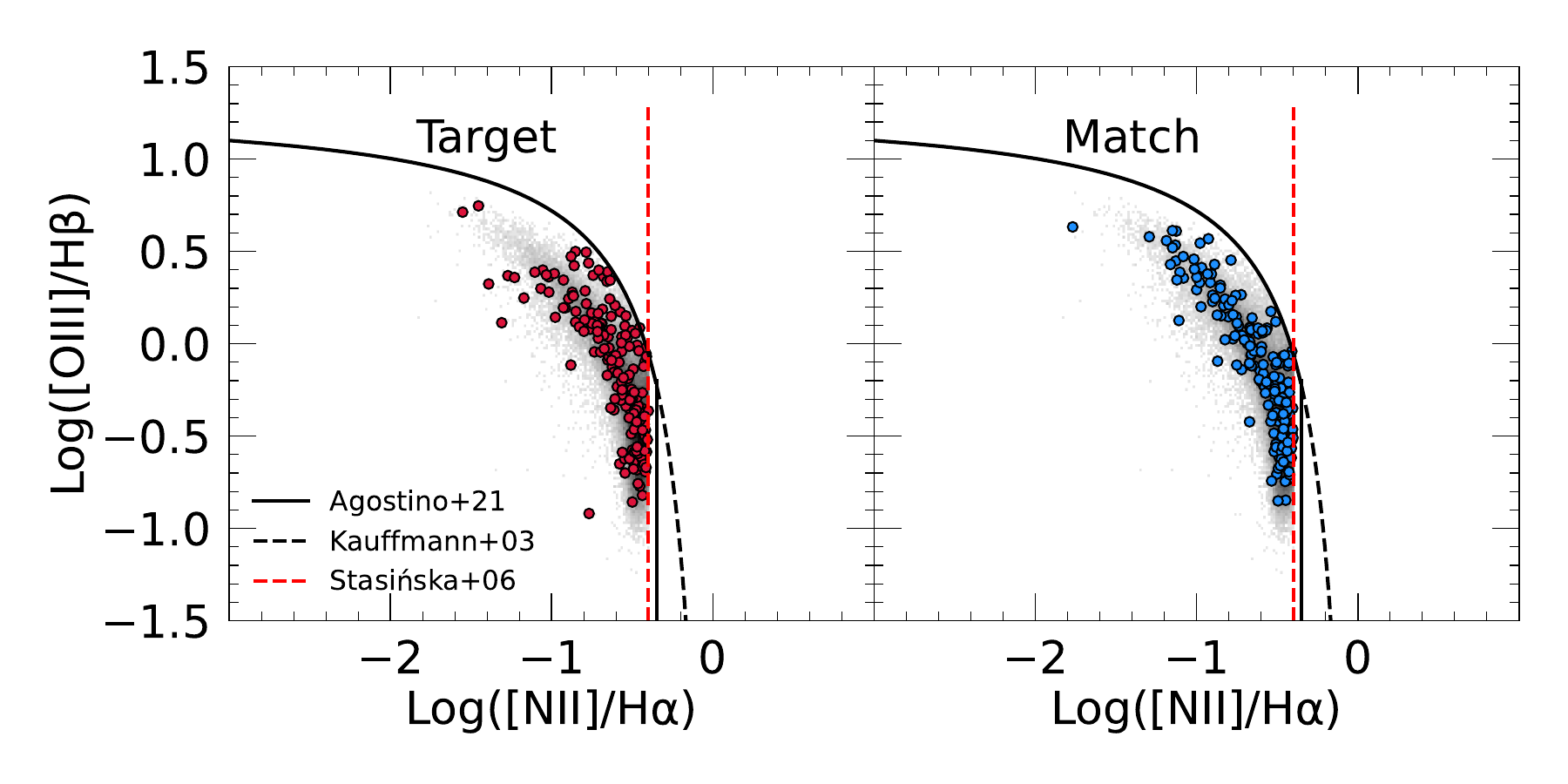} 
\caption{BPT comparison of very red and matched blue SF galaxies. \textit{Left}: BPT diagram for red star-forming (SF) galaxies with $(W1-W2)_{[z=0]}>0.8$. A subset of these galaxies appear to be offset from the ridge of SF sequence and closer to the \citet{Kauffmann2003} demarcation line. \textit{Right}: BPT diagram for blue SF galaxies ($(W1-W2)_{[z=0]}<0.5$) matched in stellar mass, SFR and redshift to red galaxies. The matched galaxies closely follow the SF ridge, indicating that the offset seen in some red SF galaxies is not driven solely by global host properties. The dashed black curve shows the \citet{Kauffmann2003} demarcation, the red dashed line shows the \citet{Stasinska_2006} vertical cut and the solid black curve shows the modified \citet{Agostino_2021} demarcation. }
\label{fig:matching}
\end{figure}

\subsubsection{Are red SF galaxies AGNs despite their BPT classification status?} \label{subsec:res_sf_color}

A red $W1-W2$ color is the consequence of the presence of warm dust. Although it is not expected that stars can achieve such heating in the absence of AGN, the observational evidence so far has been mixed, as discussed in Section \ref{sec:intro}. To investigate whether intense star formation can drive red mid-IR colors, we examine the relationship between SFR and $W1-W2$ color for our spectroscopically classified sample, as shown in the top left panel of Figure \ref{fig:drivers2}. Red SF galaxies span a wide range of SFRs (0.1 to 100 $\mathrm{M_\odot \ yr^{-1}}$) and although higher SFR correlates with red colors across all galaxy types, the large majority of high-SFR SF galaxies are not red. Thus, SFR alone cannot explain the red colors observed in our SF population. 

The top right panel of Figure \ref{fig:drivers2} shows the SFR normalized by stellar mass (the specific star formation rate (sSFR)). Again, red SF galaxies are found to occupy a wide range of sSFRs, and again they are not preferentially found among the higher sSFRs. We have performed the same analysis using the distance from the star-forming main sequence, which is a more robust tracer of burstiness than sSFR \citep{Salim2023ApJ...958..183S}, but the results remain qualitatively the same: starbursts cannot explain red colors (plots not shown). We reach similar conclusions in regards to the gas-phase metallicity --- it does not explain red colors either, in agreement with \citet{shobita2014ApJ...784..113S} and \citet{Connor2016MNRAS.463..811O}. In Section \ref{sec:discussion} we will present additional, modeling-based evidence that red $W1-W2$ color is incompatible with stellar-heated dust.

Similarly, in the upper middle panel of Figure \ref{fig:drivers1} we see a slight positive correlation between global dust attenuation and $W1-W2$ for the blue SF galaxies, but red galaxies do not preferentially have high attenuation. This again suggests a disconnect between the stellar populations and the cause of red colors.

If stars are not responsible for the red $W1-W2$ colors of BPT SF galaxies, that leaves AGN as the only option. However, it would be good to have some additional evidence for it. The top right panel of Figure \ref{fig:drivers1} shows that for SF galaxies the mid-IR excess remains on average near zero for blue galaxies. This is exactly what we expect when no AGN is present: the SFR (or equivalently, the total IR luminosity) inferred from the mid-IR flux (typically W4 at 22 $\micro$m) matches the SFR (or equivalently, the luminosity absorbed by the dust) obtained from the UV/optical SED fitting. However, the excess becomes predominantly positive as soon as we cross the $(W1-W2)_{[z=0]}$ boundary. While this excess is much less compared to red Sy2 galaxies and LINERs, it does suggest that some non-stellar heating source is responsible for red color. 

BPT classification is binary --- a galaxy either is, or is not classified as SF or AGN. We next test if there might be some more subtle differences between blue and red SF galaxies in that respect. In the left panel of Figure \ref{fig:matching}, we again show SF galaxies in our sample, but now we overplot SF galaxies with very red WISE color ($(W1-W2)_{[z=0]} >0.8$). Interestingly, a fraction of these red galaxies do not follow the ridge of the branch, but are offset towards the AGN demarcation line (similar to what was seen in \citealt{sartori2015MNRAS.454.3722S}). To see if this offset might be AGN related, or is driven by other factors, such as the metallicity or ionization parameter, \citealt{Hainline2016ApJ...832..119H}), we compare these very red SF galaxies ($(W1-W2)_{[z=0]} >0.8$) with blue ($(W1-W2)_{[z=0]} < 0.5$) galaxies, also of SF class, i.e., the galaxies where we definitely do not expect an AGN. Our goal is to isolate the effect of galaxy properties like $SFR$ and $M_*$ on WISE colors to see if red SF galaxies might have a different location in the BPT diagram. We implement a least-distance matching in the multidimensional parameter space of SFR, $M_*$ and $z$:
\[
D = \sqrt{(\Delta \log \ M_*)^2 + (\Delta \log \ \mathrm{SFR})^2 + 2(\Delta \log \ z)^2 }
\]
\noindent and for each of 181 red galaxies select the blue SF galaxy that minimizes $D$.

Matched galaxies can be seen in the right panel of Figure \ref{fig:matching}. They more closely follow the ridge of the SF branch than the red galaxies. Altogether, this analysis suggests that at least in some of the red SF galaxies the BPT hints at an AGN contribution to emission lines. However, a majority of red galaxies are squarely in the SF region.

\begin{deluxetable*}{cccccccc}

\tablecaption{BPT Star-Forming Galaxies with $(W1-W2)_{[z=0]}>0.5$ (mid-IR AGNs).
\label{table:red_sf_w1w2}}
\tablehead{
\colhead{SDSS ObjID} &
\colhead{RA} &
\colhead{Decl.} &
\colhead{Redshift} &
\colhead{log $M_*$} &
\colhead{log $M_{\mathrm{bulge}}$} &
\colhead{$(W1-W2)_{[z=0]}$} &
\colhead{$\sigma$} \\
\colhead{} &
\colhead{(deg)} &
\colhead{(deg)} &
\colhead{} &
\colhead{($M_\odot$)} &
\colhead{($M_\odot$)} &
\colhead{(mag)} &
\colhead{(km s$^{-1}$)}
}
\startdata
1237657190369132776 & 1.850906 & $-0.511188$ & 0.073 & 10.024 & 9.185 & 0.738 & 43.3 \\
1237657190369198314 & 1.998659 & $-0.589300$ & 0.187 & 11.159 & 10.942 & 1.800 & 69.9 \\
1237663783124533428 & 2.055386 & $-0.653890$ & 0.073 & 9.928 & 9.586 & 0.957 & 50.0 \\
1237657190369329237 & 2.199417 & $-0.625105$ & 0.040 & 10.089 & 9.623 & 0.848 & 50.8 \\
1237663783124664449 & 2.275286 & $-0.634965$ & 0.073 & 10.139 & 9.103 & 0.773 & 66.0 \\
1237663783124664462 & 2.312767 & $-0.699487$& 0.108 & 10.689 & 10.172 & 0.558 & 78.7 \\
1237657190369394863 & 2.414728 & $-0.540164$ & 0.041 & 9.675 & 9.001 & 0.986 & 46.4 \\
1237657190369460266 & 2.557658 & $-0.540649$ & 0.095 & 10.067 & 10.047 & 0.541 & 71.5 \\
1237653652911292563 & 6.187437 & 15.792687 & 0.182 & 10.826 & 10.583 & 0.504 & 105.7 \\
1237653652911358154 & 6.229563 & 15.691887 & 0.185 & 10.867 & 10.120 & 0.532 & 110.5 \\
\enddata
\tablecomments{
Sample table to show the first 10 rows of the full table content. Stellar mass comes from GSWLC-M2. The bulge mass comes from \citet{mendel14}. Velocity dispersion ($\sigma$) is based on the width of the forbidden lines and comes from the MPA/jHU catalog.
}
\end{deluxetable*}

By excluding the star formation heating as the cause for red $W1-W2$ colors, and by finding evidence consistent with the cause being an AGN, we conclude that red galaxies of BPT SF class are most likely true AGN. Might such AGNs be in different types of hosts than the red BPT-AGNs (Sy2s and LINERs)? The left panels of Figure \ref{fig:drivers3} shows the stellar mass  vs.\ $(W1-W2)_{[z=0]}$ colors for our three galaxy categories. Whereas for red Sy2 and LINER galaxies the stellar mass is exclusively high ($> 10^{10} \ M_\odot$) the case is different for red SF galaxies which have a wider range of stellar masses associated with them ($10^{8}$ to $10^{11} \ M_\odot$)---a range that overlaps with Sy2/LINERs. The total stellar mass of a galaxy might not be the most relevant parameter when it comes to the AGN activity. The middle panels of Figure \ref{fig:drivers3} show the stellar mass corresponding to the bulge. For red Sy2 and LINER galaxies, the bulge mass remains high ($\gtrsim 10^{10} \ M_\odot$), indicating that most of the stellar mass in red Sy2 galaxies is inside the bulge. However, for red SF galaxies we now see that the majority have low bulge masses ($\lesssim 10^{10} \ M_\odot$). The optimal separation, determined formally by maximizing the fraction of Sy2s above it and maximizing the fraction of SF galaxies below it, lies at $\log M_{\mathrm{bulge}}=10.16$. 

Given that the bulge/disk decompositions are quite uncertain from SDSS images ($\sigma \sim 0.24$; \citealt{Osborne2024ApJ...965..161O}), we also look at the velocity dispersions (right panels of Figure \ref{fig:drivers3}) as a more precise tracer of the central potential well. The dichotomy is even stronger than in the case of $M_{\mathrm{bulge}}$ --- red Sy2 and LINER galaxies have velocity dispersions almost entirely above $100 \mathrm{\ km\, s^{-1}}$, whereas in red SF galaxies they are almost exclusively below  $100 \mathrm{\ km\, s^{-1}}$. To conclude, we seem to have the following dichotomy: \emph{mid-IR AGNs in massive bulges (and therefore more massive SMBHs) exhibit emission-line ratios typical of AGNs (BPT-AGNs), whereas those in lower-mass bulges (which would include bulgeless galaxies) do not}.

The galaxies that seem to contain an AGN according to the mid-IR selection, but are optically inconspicuous, i.e., manifest themselves as BPT-SF are interesting from the standpoint of a quest to find intermediate mass black holes (IMBHs). This quest has mostly focused on low-mass galaxies and bulgeless galaxies, but we have shown that they might be more widely distributed among the galaxies with low bulge mass (lower velocity dispersion), even if stellar mass is high. Thus in Table \ref{table:red_sf_w1w2} we provide the listing of all 440 red ($(W1-W2)_{[z=0]}>0.5$) BPT-SF galaxies.

\begin{figure*}[htbp]
  \centering
    \begin{minipage}{0.32\textwidth}
        \centering
        \includegraphics[width=\linewidth]{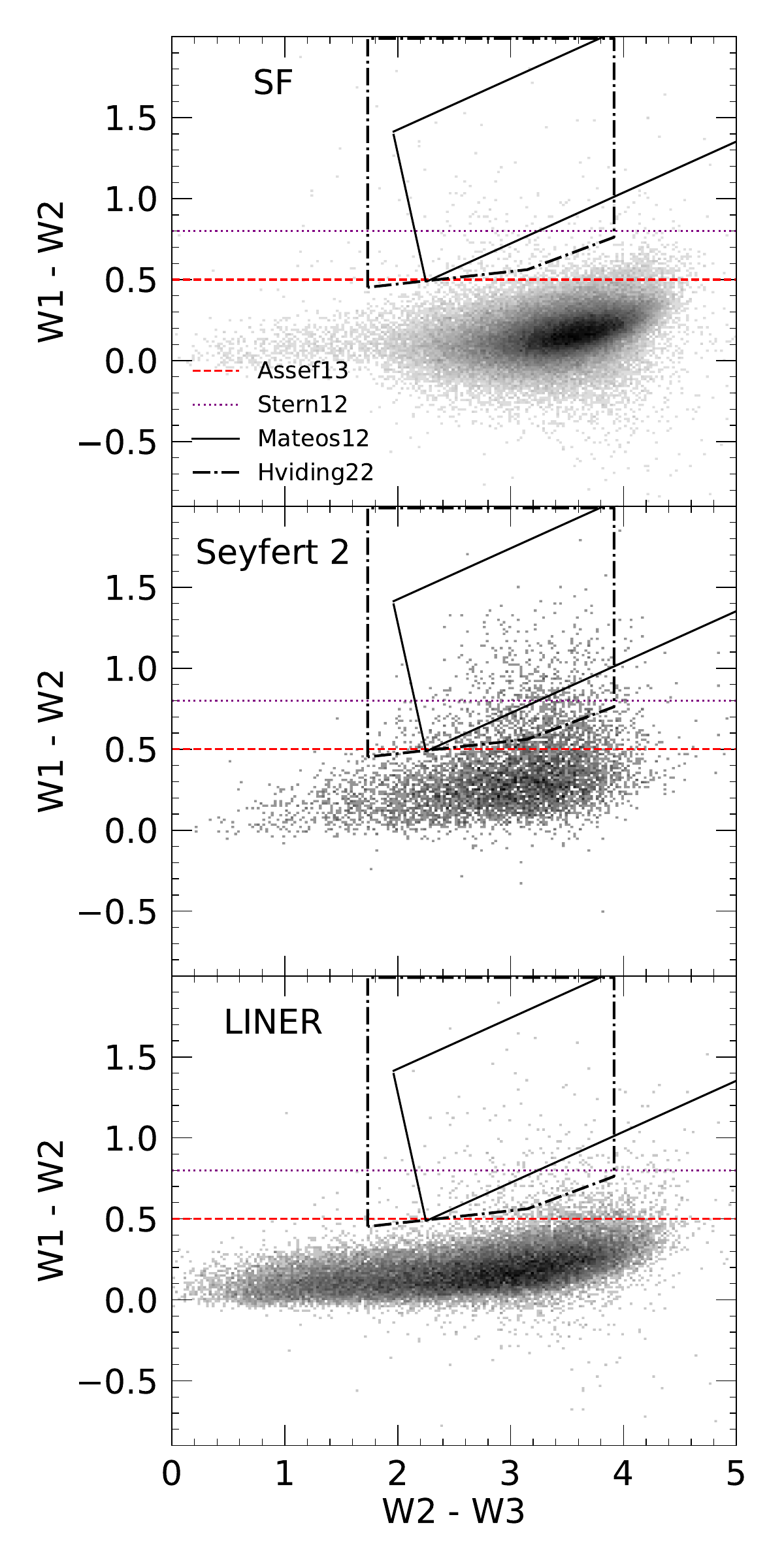}
    \end{minipage}
    \begin{minipage}{0.32\textwidth}
        \centering
        \includegraphics[width=\linewidth]{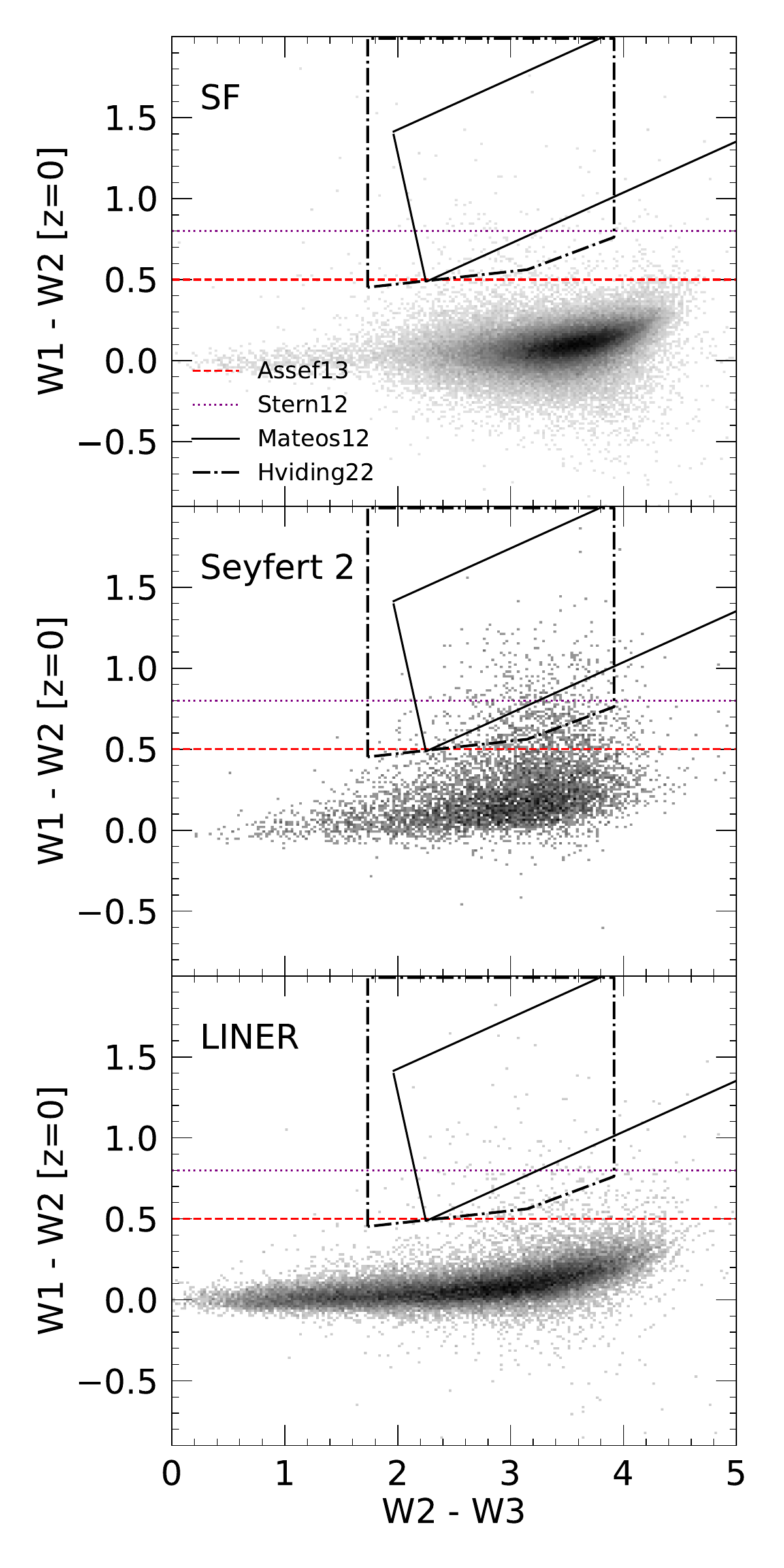}
      \end{minipage}\hfill
    \caption{Mid-IR color-color distribution and AGN selection boundaries with SF, Sy2 and LINER galaxies. \textit{Left}: $W1-W2$ vs.\ $W2-W3$. Without K-correction, SF galaxies with red colors spill over the $W1-W2>0.5$ boundary. \textit{Right}: K corrected $(W1-W2)_{[z=0]}$ vs.\ $W2-W3$. The red dashed line marks the \cite{Assef2013ApJ...772...26A} AGN selection cut at $W1-W2=0.5$, the purple dotted line marks the \cite{Stern2012ApJ...753...30S} AGN selection cut at $W1-W2=0.8$, the dash wedge marks the \cite{Mateos2012MNRAS.426.3271M} AGN selection box, and the dash dotted wedge marks the \cite{Hviding2022AJ....163..224H} AGN selection box.  Our models indicate that the K-correction of $W2-W3$ is quite small for AGN contributions that produce $(W1 - W2)_{[z=0]}>0.5$, so we are not applying it here.}
\label{fig:w1w2w3}
\end{figure*}

\section{Discussion}
\label{sec:discussion}

We focus the discussion on four questions: (1) whether better color cuts exist beyond the $W1-W2>0.5$ color cut, (2) how to reconcile our low completeness of AGN recovery with those reported in literature, (3) how useful the mid-IR selection is for lower redshift samples, and (4) what may be physical reasons for why some mid-IR selected AGNs do not show optical emission lines characteristic of AGNs. 

\subsection{Comparison of the $W1-W2 > 0.5$ cut to other cuts proposed in the literature}

The two most commonly used AGN selection criteria that are based on $W1-W2$ color alone are the $W1-W2 > 0.5$ cut that we have adopted for this study and the $W1-W2 > 0.8$ cut introduced by \citet{Stern2012ApJ...753...30S}. We show these two cuts on a color-color plot (Figure \ref{fig:w1w2w3}), which includes $W2-W3$ in addition to $W1-W2$. Left columns show $W1-W2$ without the K-correction, whereas the right columns have the K-correction applied to it. Looking at the SF panel without the K-correction we see that the redder galaxies in $W2-W3$ seem to spill over the $W1-W2 = 0.5$ cut, so it might seem that a stricter cut, like the $W1-W2 = 0.8$ one, might be necessary to eliminate this likely contamination. However, in the K-corrected version of the plot, also for SF galaxies, we see that most of this spillover has been eliminated, and that the $(W1-W2)_{[z=0]} = 0.5$ cut does not seem to allow any obvious contamination. We also see that the blue sequence in the K-corrected plots is tighter for all three galaxy populations (SF, Sy2 and LINER) than in the corresponding non-corrected plots, additionally attesting to the benefits of this correction.  

In Section \ref{sec:result} we have showed that among the fixed cuts in $(W1-W2)_{[z=0]}$ the one using 0.5 is optimal, and it also has a higher completeness of  Sy2 recovery compared to redder cuts. Indeed, applying the $(W1-W2)_{[z=0]} > 0.8$ cut to our sample would recover only 3.7\% of Sy2 galaxies, compared to 13.6\% for our nominal $(W1-W2)_{[z=0]} > 0.5$ cut (Table \ref{tab:mir_selection}).

We now move to more complex cuts proposed in the literature that are not constant in  $W1-W2$ but depend on the $W2-W3$ color. We will refer to them as slanted cuts. The first slanted cut was that of \citet{Jarrett2011ApJ...735..112J}. Two more slanted cuts, from \citet{Mateos2012MNRAS.426.3271M} and \citet{Hviding2022AJ....163..224H} are shown in Figure \ref{fig:w1w2w3}. The one by \citet{Hviding2022AJ....163..224H}, which is close to constant ($W1-W2\approx 0.5$) for $W2-W3<4$ and then tilts more, was specifically designed to follow the region in color-color space with higher fraction of SDSS BPT-selected AGNs. Indeed, the SF galaxies that spill over the $W1-W2=0.5$ line (not K corrected) are mostly eliminated by this slanted cut. However, in the color-color space where $W1-W2$ is K corrected, the need for this or any other slanted cut does not appear to be very strong. Furthermore, as shown in Table \ref{tab:mir_selection}, applying the \cite{Mateos2012MNRAS.426.3271M} or \cite{Hviding2022AJ....163..224H} criteria leads to a reduction in completeness of Sy2 recovery compared to the $(W1-W2)_{[z=0]}=0.5$ cut. This outcome is expected: shifting the selection toward redder mid-IR colors preferentially removes a substantial portion of the AGN population that does not reach such extreme colors.

We conclude that once the K-corrections are applied to $W1-W2$, one of the key improvements in our analysis, much of the contamination that might have motivated some of the more complex selection boundaries is substantially diminished. 



\begin{table}[htbp!]
\caption{Counts and percentages of galaxies selected by different Mid-IR selection criteria. A12 is the \cite{Assef2013ApJ...772...26A} selection, S12 is \cite{Stern2012ApJ...753...30S} selection, M12 is the \cite{Mateos2012MNRAS.426.3271M} selection and H22 is the \cite{Hviding2022AJ....163..224H} selection. }
\hspace*{-0.09\columnwidth}
\makebox[\columnwidth][c]{%
  \resizebox{1.2\columnwidth}{!}{%
\begin{tabular}{lccccccccc}
\hline
\textbf{Dataset} & \textbf{N\tnm{a}} & \textbf{N(A13)\tnm{b}} & \%  & \textbf{N(S12)} & \%  & \textbf{N(M12)} & \%  & \textbf{N(H22)} & \%  \\
\hline
SF        & 82110 & 349  & 0.4  & 100  & 0.1  & 85   & 0.1  & 157  & 0.2 \\
Seyfert 2 & 6749 & 918  & 13.6 & 247  & 3.7  & 265  & 3.9  & 571  & 8.5 \\
LINER     & 35078 & 349  & 1.0  & 77   & 0.2  & 74   & 0.2  & 151  & 0.4 \\
\hline
Total     & 123,937 & 1616 & 1.3 & 424 & 0.3 & 424 & 0.3 & 879 & 0.7 \\
\hline
\end{tabular}
}}
\tablecomments{Table note a: The total selected values for SF, Sy2 and LINER galaxies differ from the numbers provided in Table \ref{table:sample B} due to requiring valid entries for W3 band which was not previously accounted for in the two band selection in Table \ref{table:sample B}. Table note b: Similar to table note a. }

\label{tab:mir_selection}
\end{table}

\subsection{Reconciling various estimates of AGN completeness}

While previous studies have pointed out that the mid-IR selection will not select AGNs that are not bolometrically dominant in the IR, and our modeling exercise in Section \ref{sec:modeling} confirms that by showing that only systems with high AGN contribution to IR luminosity ($\texttt{fracAGN}\gtrsim 0.75$) achieve $W1-W2$ color redder than 0.5, the prevalence of AGNs which do not produce sufficient mid-IR emission relative to their hosts to fulfill these selection limits has not been fully quantified.  

While some previous studies report mid-IR AGN completeness values of $50-95\%$, our completeness for Sy2 (and even more so LINER) galaxies is dramatically lower (13\% and 1\%, respectively). This discrepancy reflects a key difference in how the completeness is defined among different studies. Mid-IR completeness estimates were often measured relative to AGN subsets which are IR bright. Examples include \cite{Stern2012ApJ...753...30S}, which used luminous quasars in the COSMOS field, \cite{Assef2013ApJ...772...26A}, which used AGNs selected from deep Spitzer IRAC imaging or \cite{Mateos2012MNRAS.426.3271M}, which used X-ray luminous AGNs with $L_X> 10^{44}\ \mathrm{erg\, s^{-1}}$. We calculate completeness relative to the optically (BPT) selected AGNs, which being a more sensitive tracer of AGN activity, leads to lower completeness. Our completeness is most similar to that of \cite{Hviding2022AJ....163..224H}, who evaluate it against spectroscopically classified AGNs (though, unlike us, both Type 1 and Type 2).

\begin{figure}[ht!]
\centering
\includegraphics[width=0.45\textwidth]{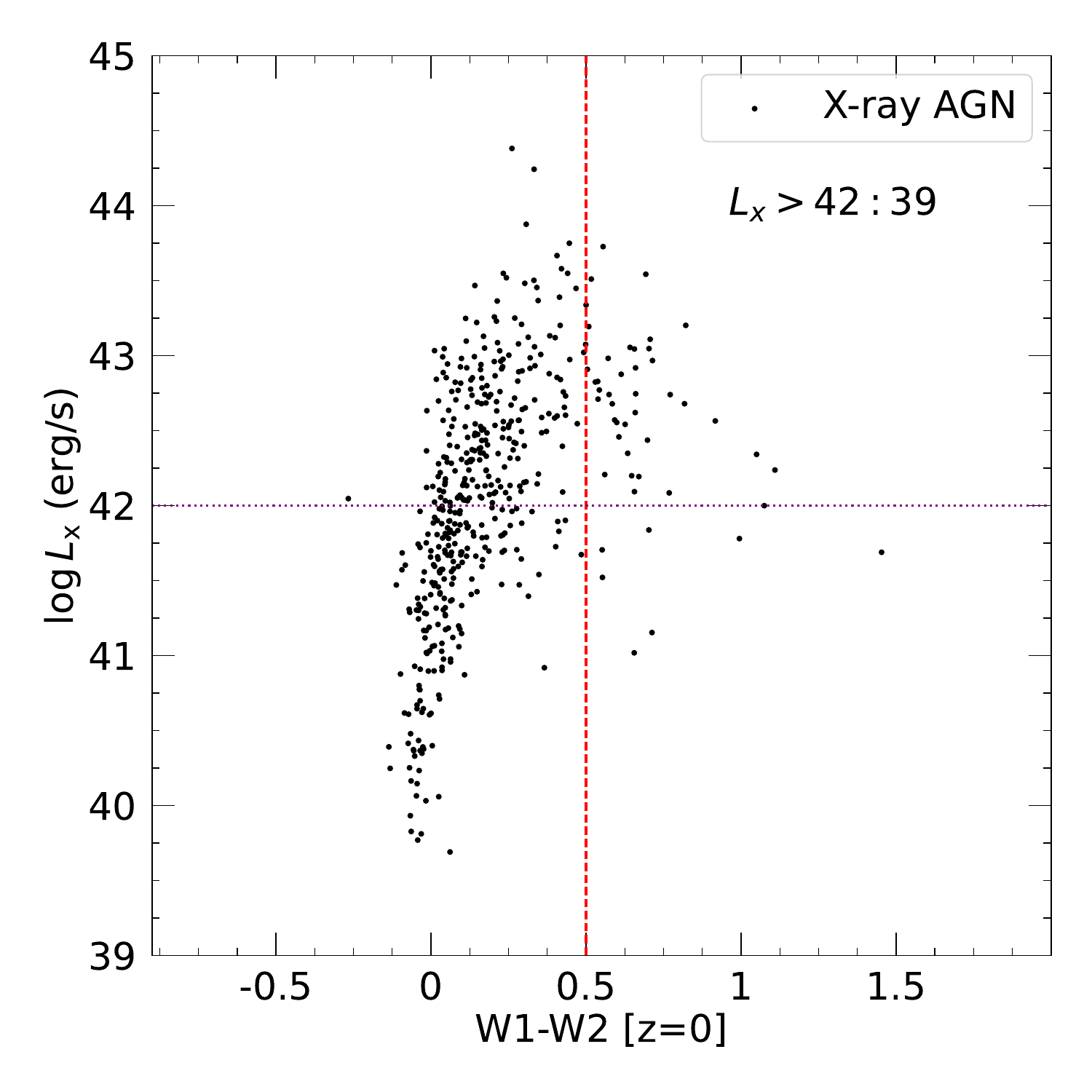} 
\caption{X-ray luminosity ($L_X$) as a function of K-corrected WISE color $(W1-W2)_{[z=0]}$ for X-ray selected AGN. The vertical dashed red line marks the $(W1-W2)_{[z=0]} = 0.5$ color cut, while the horizontal dotted purple line indicates the X-ray luminosity threshold at $\log L_X > 42$ erg s$^{-1}$, often used to separate strong AGNs. Only 15\% of X-ray AGNs in our $z<0.3$ sample satisfy the mid-IR color cut.}
\label{fig:xray}
\end{figure}

\citet{lamassa2019ApJ...876...50L} showed that the mid-IR and X-ray AGN selection are  complementary: 61\% of X-ray AGN are missed by mid-IR criteria, while 58\% of mid-IR AGNs lack X-ray detections. The overlap between X-ray- and mid-IR–selected AGNs is known to be strongly luminosity-dependent (e.g., \citealt{Eckart2010ApJ...708..584E}), with mid-IR selection favoring luminous quasars and X-ray selection sensitive also to lower-luminosity, Seyfert-like AGNs (\cite{Assef2018ApJS..234...23A}). Motivated by these studies, here we again look at the completeness of our $z<0.3$ sample and $(W1-W2)_{[z=0]}>0.5$ selection, but now against the X-ray AGNs selected from XMM-Newton catalog. We find that 73 out of 473 X-ray-selected AGNs in our sample (15\%) exhibit $(W1-W2)_{[z=0]}>0.5$, as shown in Figure \ref{fig:xray}. This relatively low completeness (but similar to that we found for Sy2s) is because our $z<0.3$ sample contains many AGNs with lower X-ray luminosities. Even when considering only the AGNs with $L_X > 10^{42}$ erg s$^{-1}$, still only 64 out of 240 (25\%) satisfy the color cut, in agreement with the conclusions of \citet{Donley2007ApJ...660..167D} or \citet{lamassa2019ApJ...876...50L}. Thus, we see that, as in the case of Sy2 and [OIII] luminosity (Figure \ref{fig:drivers1}), high X-ray luminosity might be a necessary, but is not a sufficient condition for red $W1-W2$ color.

Taken together, our work indicates that in addition to low-luminosity AGNs, which, as expected, fail to reach the $W1-W2$ color threshold, even for the high-luminosity AGNs it is difficult to tell which will be red in $W1-W2$, since neither the X-ray luminosity nor the [OIII] luminosity, or the Eddington ratio derived from the latter, correlates very strongly with the $W1-W2$ color. The relevant properties may operate on scales of dust torii ($\sim 0.1$ pc, \citealt{Chen2023MNRAS.522.3439C}).

\subsection{Benefits of mid-IR AGN selection at lower redshifts}

Given the limited completeness of mid-IR selection for low redshift AGN population, an important question is how useful is the mid-IR selection, especially if high quality optical spectroscopy is available. Our analysis suggests that the answer depends on the physical regime of the AGN and what we mean by useful.

Mid-IR selection was originally developed as a fairly efficient way to identify AGNs across a wide range of redshifts, in samples that often lacked spectra. When emission-line diagnostics are available and yield a clear AGN classification, as is the case for SDSS at lower redshifts, the mid-IR selection might not add much to complete the census of high-luminosity AGN. More precisely, in our sample we see that it is the AGNs with high velocity dispersions (or equivalently, \textit{higher bulge masses}), for which the mid-IR simply confirms what the emission lines already reveal. If mid-IR selected objects are to be admitted nonetheless, especially using a $W1-W2=0.5$ cut, we note that it is important that the $W1-W2$ color be K corrected in order to avoid introducing non-AGN contamination. 

However, even for high-bulge mass AGNs there are benefits of considering the mid-IR color. First is the \textit{characterization} of AGNs: galaxies whose SEDs reach redder $W1-W2$ colors correspond to galaxies with high AGN fractions in the IR and may correspond to specific stage in AGN triggering or evolution \citep{Ellison2025OJAp....8E..12E}. Second concerns LINERs, which are often not included among the AGNs out of concern that they are not true AGNs. However, as we have shown (lower left panel of Figure \ref{fig:drivers1}), there is little doubt that they are AGNs when their $W1-W2$ is red. Similar rationale applies to objects that cannot be classified on the BPT diagram because some of the lines are weak. As a matter of fact, the pie chart in Figure \ref{fig:piechart} shows that there are as many LINERs and what we call ``Additional LINERs or Sy2s" (where either [OIII] or H$\beta$ is too weak for BPT classification (SNR$<2$)) combined, as there are confirmed Seyfert 2s among the red objects. Thus, mid-IR selected AGNs can significantly expand the AGN samples available for study.

Situation is more complex in cases where the K-corrected $W1-W2$ is red, but the galaxy falls in the SF (i.e., non-AGN) region of the BPT diagram. Interestingly, we have seen that such galaxies typically have lower bulge masses ($<10^{10} M_{\odot}$), or equivalently, lower velocity dispersions ($<100$ km s$^{-1}$), and it is not at all inconceivable that the BPT diagram might fail to identify an obscured AGN in such cases. However, as many studies have pointed out, one needs to make sure that the red $W1-W2$ color is not the result of star formation heated dust, rather than an AGN. 

Our analysis in Section \ref{subsec:res_sf_color} suggests that galaxies with SF BPT classification, but with $(W1-W2)_{[z=0]}>0.5$, are indeed AGNs. From the observational standpoint, we find that these red SF galaxies exhibit a mid-IR excess in $W4$ (22 $\micro$m). This excess basically measures how much more W4 luminosity is present compared to what is expected from SF alone. Importantly, this excess is measured without $W1$ or $W2$. The excess is present for most Sy2s, whether blue or red in $W1-W2$, but only for red SF. Blue SF galaxies do not have it, suggesting it is a meaningful measure of AGN presence. In addition to this, we have also seen that at least some of the red SF galaxies lie closer to the AGN demarcation in the BPT diagram than their blue doppelgangers (Figure \ref{fig:matching}), suggesting a subtle emission-line signature of an AGN.

\begin{figure}[ht!]
\centering
\includegraphics[width=0.45\textwidth]{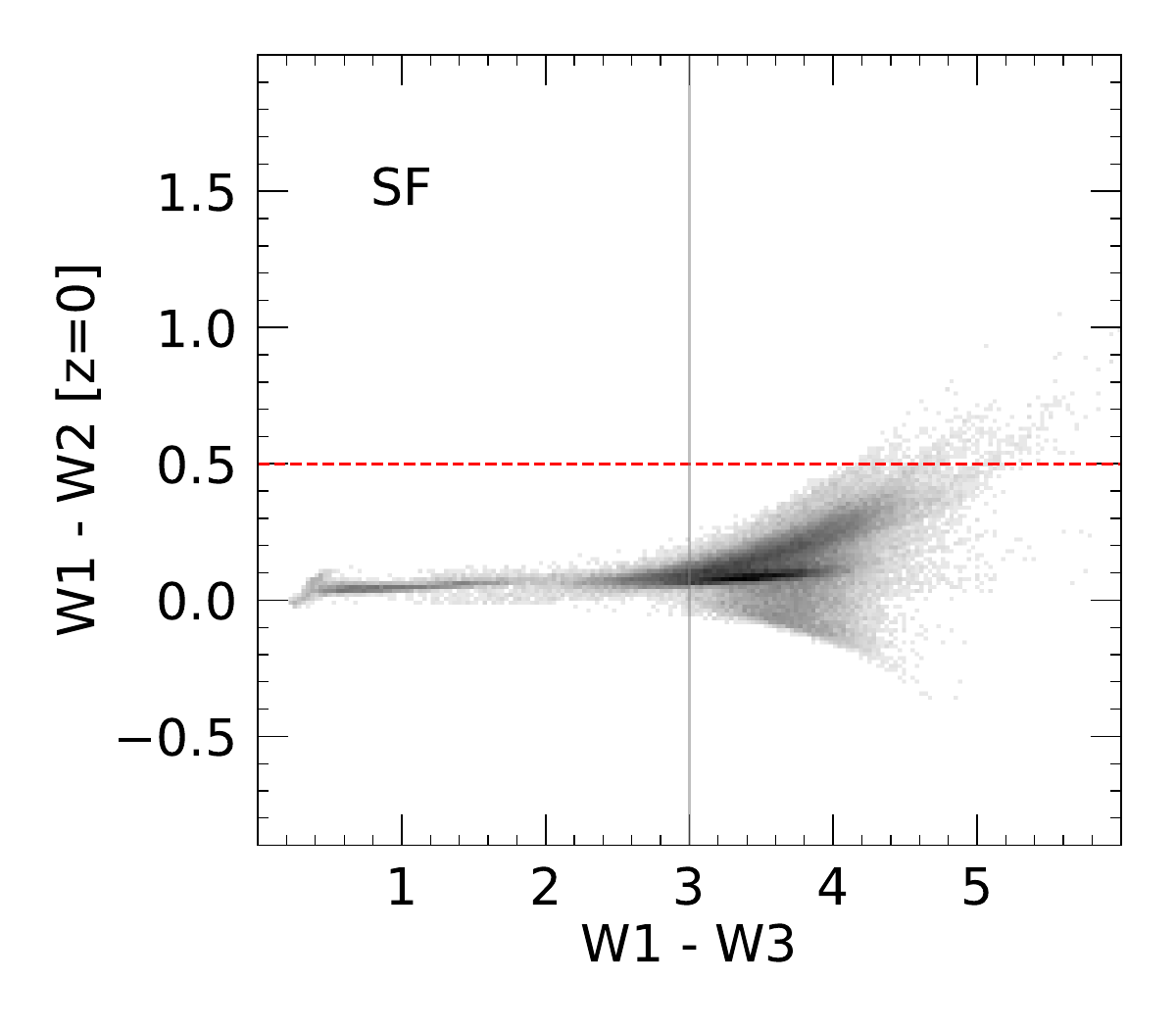} 
\caption{ Predicted WISE colors from star-formation-heated dust models for SF galaxies. $(W1-W2)_{[z=0]}$ and $W2-W3$ colors for SF galaxies obtained with CIGALE  using \cite{2007ApJ...657..810D} dust models (updated in \citealt{2014ApJ...780..172D})). The models were constrained using only $W3$ and $W4$ fluxes so that $W1-W2$ is effectively a prediction using stellar heating alone. No combination of star-formation-heated dust parameters reproduce the observed colors (Figure \ref{fig:w1w2w3} upper right panel), indicating that an additional AGN-heated dust component is required to explain the population of red SF galaxies.}
\label{fig:dust models}
\end{figure}

From the modeling standpoint, here we present a result that also supports the interpretation that red SF galaxies are AGNs, or more precisely, that they cannot be sufficiently heated by SF alone to explain their red colors. The key to this modeling exercise lies in the use of \cite{2007ApJ...657..810D} dust templates (updated in \citealt{2014ApJ...780..172D}), which allow for a wide range of intense radiation fields, but still not the kinds that would be produced by AGNs. We fit these models using CIGALE, allowing for all possible values for the PAH mass fraction, minimum radiation field and the radiation field power slope exponent. Importantly, we fit the models only to $W3$ and $W4$ fluxes, because we wish the models to in essence predict what the $W1-W2$ color would be if only SF heating was assumed. The results are shown in a color-color plot in Figure \ref{fig:dust models}, which should be compared to the upper right panel of Figure \ref{fig:w1w2w3} (the actual colors of red SF galaxies). We find that even when the models attain red $W1-W2$ colors ($(W1-W2)_{[z=0]}>0.5$), these rarely exceed 0.7, and at most reach $\sim 1.0$, whereas in reality there are many more red SF galaxies with $(W1-W2)_{[z=0]}>0.5$, and some reaching up to $\sim 1.5$. Secondly, even when model colors exceed $(W1-W2)_{[z=0]}=0.5$, the corresponding $W2-W3$ colors do not match the observed colors of SF galaxies. Model $W2-W3$ colors are always only red ($W2-W3>4$), whereas the observed colors span a wide range ($1<W2-W3<4$). That SF heated dust is generally expected to produce red $W2-W3>4$ when $W1-W2$ is also red, has also been shown in modeling by \citet{Hainline2016ApJ...832..119H} and \citet{Shobita2018ApJ...858...38S}. 

To conclude, no combination of parameters for star-formation heated dust in the \cite{2007ApJ...657..810D} framework reproduces the WISE colors seen in red WISE SF galaxies. Combined with our analysis in Section \ref{subsec:redwise}, this provides evidence for the argument that the observed red colors in red SF galaxies require an AGN-heated dust component. This does not mean that strong starbursts may not also be present in some red SF galaxies, but ultimately the red $W1-W2$ color seems to be driven by an AGN. A dramatic example of such scenario is the red SF dwarf recently studied using JWST in \citet{Doan2025ApJ...987...99D}, where nuclear starburst is present, but ultimately the red mid-IR continuum is associated with an unresolved, very small ($<5$ pc) nuclear source. 

We conclude that even though it identifies only a small fraction of optically defined AGN, the mid-IR selection is able to identify important categories of AGNs that would be missed or not included by optical selection, specifically the AGNs in low bulge mass hosts, AGNs in LINERs and AGNs in galaxies with weak emission lines that are not subject to BPT selection. And by using the K-corrected $W1-W2$ this is achievable with minimal non-AGN contamination. 

\subsection{AGNs in hosts with low bulge mass}

Here we attempt to derive a physical picture for galaxies that appear to be AGNs based on the mid-IR ($W1-W2$) color, but present themselves as non-AGNs based on the emission-line diagnostics. Previously, the population of such inconsistently classified AGNs was highlighted among the galaxies with bulge-to-total ratios consistent with zero (bulgeless galaxies, \citealt{shobita2014ApJ...784..113S}), which have moderately high stellar masses spanning the range $10.0 < \log  M_{*} < 11.3$, and among the dwarf galaxies \citep{sartori2015MNRAS.454.3722S,Hainline2016ApJ...832..119H}, selected to have $\log  M_{*} < 9.5$. We see (upper left panel of Figure \ref{fig:drivers3}) that while red SF make a higher fraction among the lower-mass and dwarf galaxies (as found by  \citealt{shobita2014ApJ...784..113S,Connor2016MNRAS.463..811O}), they span a full range of stellar masses ($8 < \log  M_{*} < 11.5$) that entirely encompasses the mass range of Seyfert 2s. Therefore, the total stellar mass of a galaxy is a poor predictor of whether the mid-IR selected AGN will appear as an emission-line AGN or not. The distinction, however, is more pronounced when we consider the \textit{bulge stellar mass} (middle panels of Figure \ref{fig:drivers3}), which shows that regardless of their total mass, the bulge masses of red SF galaxies tend to be $\log M_{\mathrm{bulge}} \lesssim 10$, whereas the bulge masses of BPT AGNs tend to be $\log M_{\mathrm{bulge}} \gtrsim 10$. The dichotomy is even more pronounced in  \textit{velocity dispersion}, which might serve as a more robust indicator of the central mass than the photometrically determined bulge-to-total ratio \citep{Osborne2024ApJ...965..161O}, and appears to be fundamentally more strongly correlated with the central black hole mass than the bulge mass \citep{beifiori12,huber25}. There we see that most mid-IR AGNs that lie in the SF region of the BPT have $\sigma < 100$ km s$^{-1}$, whereas the ones classified as BPT AGNs overwhelmingly have $\sigma > 100$ km s$^{-1}$.

What physical processes might give rise to this dichotomy? One often invoked explanation for why some robust AGNs are ``misclassified" as SF in the BPT diagram is the dilution of AGN lines by SF present within the spectroscopic fiber. However, although red SF galaxies tend, on average, to have higher sSFRs than red Sy2s and LINERs (Figure \ref{fig:drivers2}, right panels), the range of sSFRs also overlaps between BPT AGNs and non-AGNs, suggesting that other factors may be involved. Another explanation is that the AGN lines are weak due to the nuclear dust attenuation, which would make the attenuated AGN lines more amenable to SF dilution. While this is certainly reasonable given that these are mid-IR-selected AGNs where obscuration must be important, it is less clear why this would affect only the AGNs in low bulge masses. Instead, we propose an alternative explanation in which the AGN lines are \textit{intrinsically} weak in low-bulge mass galaxies. Indeed, in Figure \ref{fig:drivers1} (left panels) we see that the [OIII] luminosity of red SF galaxies tends to be significantly lower (at least 10 times) than of BPT AGNs. Given that the [OIII] also comes from HII regions, the true discrepancy might even be higher. We can think of two reasons for why the lines might be stronger in massive bulges but weaker in less massive ones:

\begin{enumerate}
\item The more massive bulges might have resulted from more intense interactions (like minor mergers), which in turn might increase the covering fraction of NLR gas, as well as the dust around the nucleus, raising the probability that the AGN radiation will be re-processed into NLR-type emission and the mid-IR.

\item We expect the central BHs of galaxies of lower bulge mass to also to be of lower mass, which would result in less luminous AGN and weaker lines.
\end{enumerate}

Another possibility, especially relevant for the bulges of lowest masses, is that the accretion onto IMBHs intrinsically produces line emission with ratios similar to those of HII regions, as shown in the models of \citet{cann2019ApJ...870L...2C}. The connection with the bulge mass (velocity dispersion), and therefore potentially the BH mass, is why we consider the intrinsically weak lines or intrinsically HII-like line ratios to be the more likely explanation for the observed dichotomy than the straightforward dilution or dust obscuration.

To conclude, the fact that the mass of the central black hole is more closely correlated with the bulge mass (and even more so with the velocity dispersion) than the total stellar mass of a galaxy (e.g., \citealt{beifiori12,marsden20,huber25,shankar25}) might provide a natural explanation for why the split between the mid-IR AGNs that show emission-line signatures of AGNs and the ones that do not follows the bulge mass (or the velocity dispersion) rather than the total stellar mass of a galaxy.

\section{Conclusions}

Mid-Infrared color selection has been used as a practical tool to identify AGNs, but its reliability and completeness for the general low-redshift ($z<0.3$) galaxy population remains uncertain. Combining the large spectroscopically classified sample from SDSS (GSWLC-M catalog) with mid-IR photometry from WISE, we revisit some key questions central to mid-IR AGN identification. In particular, we focused on a simple cut in $W1-W2$, and the three galaxy catgories according to their emission lines: star-forming galaxies (BPT-SF), Seyfert 2 galaxies, and LINER galaxies. Our key findings are summarized below:

\begin{enumerate}
    \item The observed-frame $W1-W2$ color is subject to a K-correction of up to 0.2 mag at $z<0.3$. We provide simple formulae for K-correction up to $z=0.3$ (Equation \ref{eq:kcorr}).
    \item Applying the K-correction reduces the number of BPT-SF galaxies that exceed $W1-W2=0.5$ by a factor of 2.5. The removed galaxies are most likely non-AGN contaminants.
     \item Applying the K-correction removes the tail of non-AGNs with $W2-W3>4$, which would otherwise (without the K-correction) exceed the  $W1-W2=0.5$ cut. Thus, the application of the K-correction diminishes the need for slanted (i.e., $W2-W3$ dependent) $W1-W2$ selection cuts.
    \item Varying the rest-frame $W1-W2$ color cut to be bluer than 0.5 brings back a non-AGN contamination, whereas making it redder reduces the completeness with respect to Sy2s. We find that the K-corrected $W1-W2=0.5$ cut is as optimal for selecting mid-IR AGNs as it can get.
    \item The rest-frame $W1-W2>0.5$ color cut selects 13\% of Sy2s and 1\% of LINERs, underscoring that the large majority of emission-line identified AGNs fall below even this lenient mid-IR threshold. This incompleteness has previously been pointed out for dwarf galaxies \citep{Hainline2016ApJ...832..119H}.
    \item Our modeling with \citet{2014ApJ...784...83D} IR templates shows that the red color (K-corrected $W1-W2>0.5$) is achieved only when the AGN contributes $\sim 75\%$ or more to the total IR luminosity, in line with the estimates from previous studies \citep{Assef2013ApJ...772...26A}.
    \item{High AGN luminosity is a necessary, but not a sufficient condition for Seyfert 2s or LINERs to be selected as mid-IR AGNs. No parameter or a combination of parameters (e.g., the Eddington ratio) that we have investigated can predict which AGNs get to be mid-IR selected and which do not. This suggests the role of processes or physical conditions on much smaller spatial scales (dust torii sizes?)}
    \item{Seyfert 2s on average show a clear excess of $W4$ emission with respect to what is expected from SF alone (the ``mid-IR excess"). This excess is  on average present even in blue Seyferts, which the color cut does not select (K-corrected $W1-W2<0.5$) but is stronger (as expected) in red ones (K-corrected $W1-W2>0.5$)}
    \item LINERs with red colors (K-corrected $W1-W2>0.5$) show a similar mid-IR excess as Seyfert 2s, suggesting they are true AGNs.
    \item Approximately 20\% of all red sources are classified as star-formers (non-AGN) according to the BPT diagram.
    \item Despite their non-AGN emission-line classification we believe these red BPT-SF galaxies are true AGN because:
      \begin{itemize}
    \item They also show mid-IR ($W4$) excess with respect to what is expected from SF.
    \item Some of them are offset towards the AGN demarcation in the BPT diagram in a way that cannot be explained by their stellar population properties.
    \item Even though they have high sSFRs, most of the similarly high-sSFR galaxies remain blue, disfavoring the starburst explanation for red color.
    \item These red SF galaxies have $W1-W2$ and $W2-W3$ colors incompatible with the stellar dust heating models of \citet{2014ApJ...784...83D}, in line with the results of \citet{Shobita2018ApJ...858...38S}.
      \end{itemize}
    \item We find a dichotomy, in the sense that the mid-IR AGNs in massive bulges ($M_{\mathrm{bulge}}> 10^{10} M_{\odot}$) predominantly (84 \% of cases) manifest themselves as BPT AGNs (Sy2s or LINERs), whereas those in low-mass bulges ($M_{\mathrm{bulge}}< 10^{10} M_{\odot}$) are predominantly not classified as emission-line AGN, but rather as BPT star-formers (60\%). This distinction is even stronger in velocity dispersion, but is diminished in the total stellar mass of a galaxy, where red BPT-SF galaxies span a full range of masses overlapping with those of red BPT-AGNs.
    \item We hypothesize that the origin of the BPT-AGN vs.\ BPT-SF dichotomy of mid-IR AGNs is due to the low-bulge mass AGNs having smaller central BHs, which produce AGN lines that are intrinsically weak, or because their AGN line ratios are intrinsically similar to those of the HII regions \citep{cann2019ApJ...870L...2C}. Alternatively, (or additionally), the smaller bulges may have smaller NLR covering fractions.
\end{enumerate}

Overall, we have shown that even a simple cut in $W1-W2$ is quite robust in selecting pure samples of AGNs once the appropriate K-corrections are applied. And although the mid-IR cuts select only the AGN with the highest fractional IR contribution, missing a large majority of the optically identified AGN, the unique strength of mid-IR selection appears to lie in being able to identify the AGNs in low-mass ($<10^{10} M_{\odot}$) bulges (irrespective of their total stellar mass, which can sometimes be quite high), which would otherwise be missed by the BPT diagnostics. 

\acknowledgments

This work made use of the following software packages: \texttt{astropy} \citep{astropy:2013,astropy:2018,astropy:2022}, \texttt{matplotlib} \citep{hunter2007CSE.....9...90H}, \texttt{numpy} \citep{harris2020Natur.585..357H}, \texttt{pandas} \citep{mckinney-proc-scipy-2010,pandas2024zndo..13819579T}, \texttt{python} \citep{python}, \texttt{scipy} \citep{scipy_10155614, scipy_2020}, \texttt{TOPCAT} \citep{topcat2005ASPC..347...29T}, and \texttt{tqdm} \citep{tqdm2021zndo...5517697D}.

This research has made use of the Astrophysics Data System, funded by NASA under Cooperative Agreement 80NSSC21M00561.

Software citation information aggregated using \texttt{\href{https://www.tomwagg.com/software-citation-station/}{The Software Citation Station}} \citep{Wagg2024arXiv240604405W,wagg2025zndo..17145205W}.

\bibliography{template}

\end{document}